\documentclass[preprint,12pt]{elsarticle}

\usepackage{graphicx}
\usepackage{amssymb}
\usepackage{tabularx}
\usepackage{newtxmath}
\usepackage{mathbbol}
\usepackage{soul}
\usepackage{textcomp}

\usepackage{lineno}

\usepackage{hyperref}

\newcommand{\beq}{\begin{equation}}
\newcommand{\eeq}{\end{equation}}

\newcommand{\tr}{\mbox{\rm{tr}}\,}
\newcommand{\cof}{\mbox{\rm{cof}}\,}

\def\vec   #1{\mbox{\boldmath $#1$}{}}
\def\scas  #1{\mbox{{\scriptsize{${\rm{#1}}$}}}{}}

\usepackage{comment}
\usepackage{amsmath}
\usepackage{lipsum}
\usepackage{graphics}
\usepackage[dvipsnames]{xcolor}
\usepackage{ulem}

\journal{arxiv}

\begin{document}

\begin{frontmatter}


\title{Data-driven Tissue Mechanics with Polyconvex Neural Ordinary Differential Equations}



\author{Vahidullah Tac$^{1}$, Francisco S. Costabal$^{2}$, and Adrian B. Tepole$^{1,3}$ }

\address{$^1$School of Mechanical Engineering, Purdue University, West Lafayette, IN, USA\\ $^2$Department of Mechanical and Metallurgical Engineering, School of Engineering, Institute for Biological and Medical Engineering, Schools of Engineering, Medicine and Biological Sciences, Pontificia Universidad Catolica de Chile, Santiago, Chile\\$^3$Weldon School of Biomedical Engineering, Purdue University, West Lafayette, IN, USA}

\begin{abstract}
Data-driven methods are becoming an essential part of computational mechanics due to their advantages over traditional material modeling. Deep neural networks are able to learn complex material response without the  constraints of closed-form models. However, data-driven approaches do not a priori satisfy physics-based mathematical requirements such as polyconvexity, a condition needed for the existence of minimizers for boundary value problems in elasticity. In this study, we use a novel class of neural networks, neural ordinary differential equations (N-ODEs), to develop data-driven material models that automatically satisfy polyconvexity of the strain energy. We take advantage of the properties of ordinary differential equations to create monotonic functions that approximate the derivatives of the strain energy with respect to deformation invariants. The monotonicity of the derivatives guarantees the convexity of the energy. The N-ODE material model is able to capture synthetic data generated from closed-form material models, and it outperforms conventional models when tested against experimental data on skin, a highly nonlinear and anisotropic material. We also showcase the use of the N-ODE material model in finite element simulations of reconstructive surgery. The framework is general and can be used to model a large class of materials, especially biological soft tissues. We therefore expect our methodology to further enable data-driven methods in computational mechanics.
\end{abstract}


\end{frontmatter}

\section{Introduction}\label{motiv}
The past two decades have witnessed the emergence of data-driven models for describing various  engineering systems and physical phenomena in wide ranging applications, from improving medical diagnosis \cite{Ma2017diagnosis}, to weather forecasting \cite{guhathakurta2006}, energetic material property prediction \cite{casey2020prediction}, and turbulence modeling \cite{Durasaimy2019}. Data-driven methods such as deep neural networks are becoming popular in materials science and materials engineering in applications such as fatigue life prediction \cite{lee1999fatigue}, identification of material parameters \cite{lu2020multifidelity}, systems identification \cite{wang2021variational}, and constitutive modeling \cite{Le2015,liu2020,peng2020multiscale,garikipati2020multiresolution,tac2021datadriven}.

The notion of a free energy function, and the definition of the stress by differentiation of the free energy with respect to the deformation lies at the core of nonlinear elasticity, from hyperelastic to energy dissipating materials \cite{marsden1994mathematical}. The stress, in turn, has to be a strongly elliptic function of the deformation gradient in order to satisfy the existence of traveling waves with real wave speeds \cite{kuhl2006illustration}. To satisfy this requirement and to guarantee the existence of solutions in nonlinear elastostatics, Ball concluded that the stored strain energy has to be a polyconvex function of the deformation gradient \cite{ball1976convexity}. Traditionally, material model development has relied on expressing the strain energy as an explicit and differentiable analytical function of the deformation \cite{gasser2005GOH,holzapfel2000HGO,fung1979}. The notion of polyconvexity has been an important factor in the development of these closed-form material models \cite{chagnon2015hyperelastic}. In particular, advances in modeling of soft tissue material behavior has prompted the development of nonlinear and anisotropic strain energies that satisfy this polyconvexity requirement \cite{ehret2007polyconvex}.

Selection of an appropriate expert-constructed material model for a specific material usually requires advanced expertise and significant trial and error. There are a large number of models in the literature designed to fit different families of materials. Even in specific fields such as skin mechanics there is no consensus on the choice of a material model \cite{ limbert2019skin, jor2013computational, mueller2021reliability}. Furthermore, the analytical form of the strain energy function may be too restrictive for many applications, resulting in poor prediction performance. Sensitivity with respect to parameters is another issue, for instance when exponential models are used \cite{lee2018}. 

A recent trend in material modeling has been the use of deep neural networks to describe either the strain energy or its derivatives \cite{liu2020, tac2021datadriven, leng2021, Vlassis202elastoplast}. Neural networks of sufficient complexity can be used to learn, reproduce and predict the behavior of any elastic material, which resolves many issues  of expert-constructed models. However, the convexity of the strain energy function in data-driven frameworks is usually ignored or enforced with additional loss terms \cite{liu2020, leng2021, tac2021datadriven}. Using this approach does not guarantee the convexity of the strain energy. The penalty terms will push the solution space to resemble that of a convex function in the training region. However, even if convexity checks are satisfied exactly on all the training points, there is no guarantee that convexity extends outside the training points, or beyond the boundaries of the training region. Furthermore, adding loss terms to ensure convexity adds to the nonlinearity of the loss space which makes finding the global minimum a daunting task. These loss terms also result in lengthy calculations during training and may limit the flexibility of the neural networks. Thus, there is a need for data-driven methods that automatically satisfy the polyconvexity requirements but can still capture the behavior of any material.

\begin{figure}[h!]
\centering
\includegraphics[width=\linewidth]{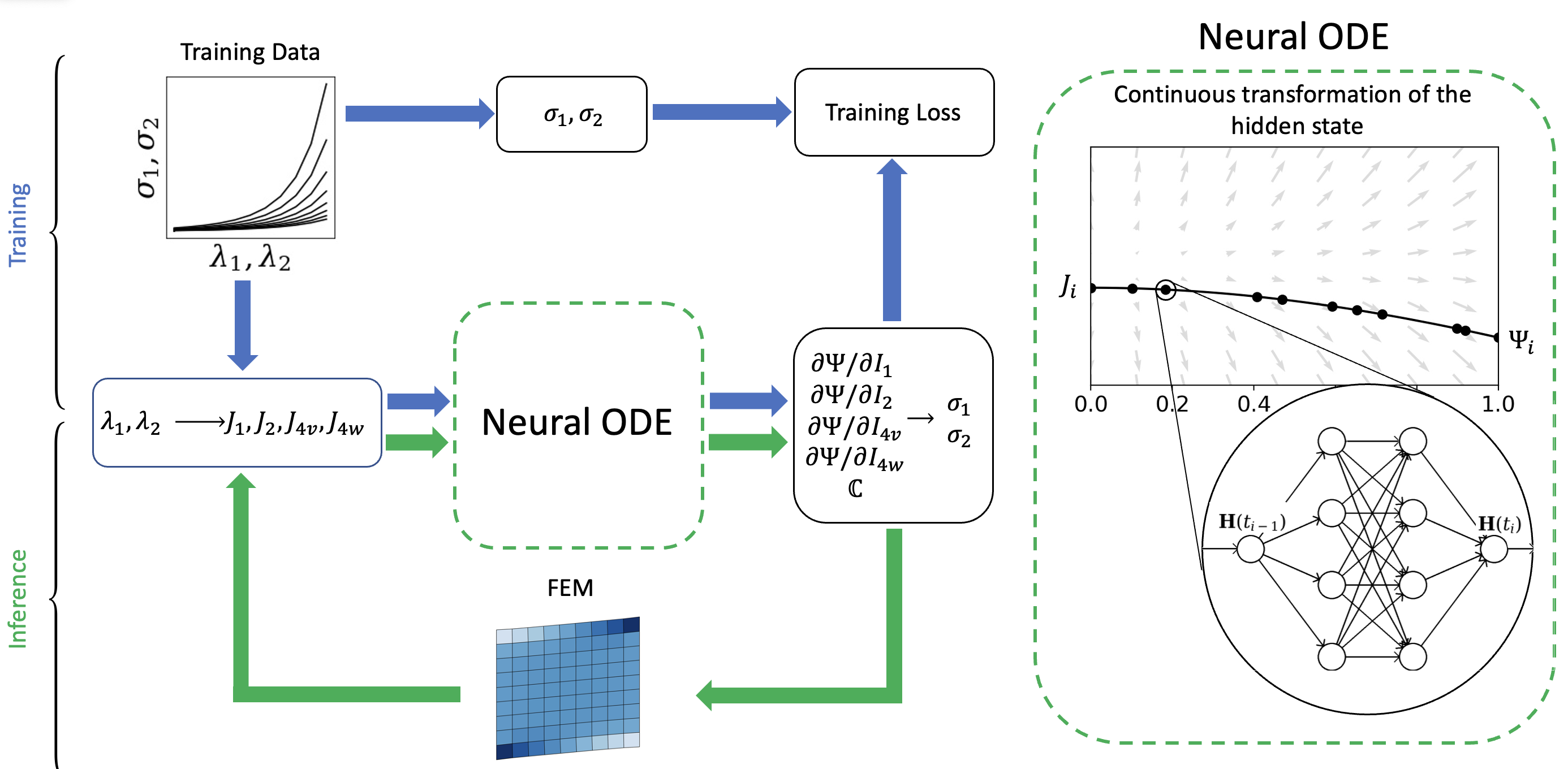}
\caption{Workflow of the training and inference processes in the N-ODE material model (left), continuous transformation of the hidden state in the N-ODE from modified invariants $\mathbf{H}(0) = J_i$, to the derivatives of the strain energy $\mathbf{H}(1) = \Psi_i$ (right).}
\label{fig_diagram} 
\end{figure}

In this study we use a new type of neural networks known as neural ordinary differential equations (N-ODE) \cite{chen2019node} to estimate the derivatives of the strain energy function with respect to invariants of deformation. Polyconvexity of the strain energy is automatically satisfied in our formulation. We train the N-ODEs using synthetic as well as experimental stress-stretch data from porcine skin biaxial experiments obtained from \cite{tac2021datadriven}. Finally, we demonstrate the applicability of the N-ODE material model in finite element simulations. The diagram in Fig. \ref{fig_diagram} shows an overview of our methodology.

\section{Materials and Methods}

\subsection{Construction of polyconvex strain energy functions}

The existence of physically realistic solutions to nonlinear elasticity problems requires polyconvexity of the strain energy function with respect to the deformation gradient $\mathbf{F}$ \cite{ehret2007polyconvex, ball1976convexity, schroder2010anisotropic}. A function $\Psi(\mathbf{F})$ is polyconvex in $\mathbf{F}$ if there is a function $\hat{\Psi}$ with  $\Psi(\mathbf{F})=\hat{\Psi}(\mathbf{F},\mathrm{cof}\mathbf{F},\mathrm{det}\mathbf{F})$ such that $\hat{\Psi}$ is convex on the extended domain formed by $\mathbf{F},\mathrm{cof}\mathbf{F},\mathrm{det}\mathbf{F}$. Constructing general functions that satisfy this requirement can be challenging, but a sufficiently flexible subset of polyconvex functions of $\mathbf{F}$ can be obtained through an additive decomposition

\begin{equation}
\Psi(\mathbf{F}) = \Psi_{\scas{F}}(\mathbf{F})+   \Psi_{\scas{cof}}(\mathrm{cof}\mathbf{F})+\Psi_{\scas{det}}(\mathrm{det}\mathbf{F})\, 
\label{eq_polyconvex_sumF}
\end{equation}

with $\Psi_{\scas{F}}$, $\Psi_{\scas{cof}}$, $\Psi_{\scas{det}}$ each a convex function. Furthermore, to guarantee objectivity of the strain energy, $\Psi$ is usually not expressed as a function of the full deformation gradient but instead of the right Cauchy Green deformation $\mathbf{C}=\mathbf{F}^\top\mathbf{F}$ . In that case the following form of the strain energy will retain polyconvexity with respect to $\mathbf{F}$,

\begin{equation}
\Psi(\mathbf{F}) = \Psi_{I_1}(I_1)+   \Psi_{I_2}(I_2) + \Psi_{J}(J)\, 
\label{eq_polyconvex_sumI}
\end{equation}

provided $\Psi_{I_1}$,$\Psi_{I_2}$ are convex non-decreasing functions of their respective arguments \cite{schroder2003invariant}, with $I_1, I_2$ the first two invariants of $\mathbf{C}$, and $J=\mathrm{det}\mathbf{F}=\sqrt{\mathrm{det}\mathbf{C}}$. The reason why constructing a function based on (\ref{eq_polyconvex_sumI}) yields polyconvex functions as expressed in (\ref{eq_polyconvex_sumF}) is because the invariants $I_1,I_2$ of $\mathbf{C}$ are convex non-decreasing functions of $\mathbf{F}$ and $\mathrm{cof}\mathbf{F}$, respectively. Explicitly, the invariants of $\mathbf{C}$, $I_1$, $I_2$,  are related to $\mathbf{F}$, $\cof\mathbf{F}$,

\begin{eqnarray}
    I_1 &=& \tr(\mathbf{C}) = \mathbf{C}:\mathbf{I},\\
    I_2 &=& \tr(\cof\mathbf{C}) = \frac{1}{2}[I_1^2 - tr(\mathbf{C}^2)].
    \label{eq_invs}
\end{eqnarray}

To consider anisotropy, additional pseudo-invariants need to be considered. A popular option is $I_{4v} = \mathbf{C}:\mathbf{V}_0$, where $\mathbf{V}_0=\mathbf{v}_0\otimes \mathbf{v}_0$ is a structural tensor defined by the anisotropy direction $\mathbf{v}_0$, which usually corresponds to a fiber orientation in soft tissue in the reference configuration. It can be shown that $I_{4v}$ is a convex function of $\mathbf{F}$. Consequently, we can include additional terms that depend on $I_{4v}$ in the strain energy function retaining polyconvexity while including the effects of anisotropy. We consider two different fiber directions $I_{4v}$ and $I_{4w}$, corresponding to the vectors $\mathbf{v}_0$ and $\mathbf{w}_0$ in the reference configuration. Lastly, we note that since $I_1$, $I_{4v}$,$I_{4w}$ are convex functions of $\mathbf{F}$, and $I_2$ is convex with respect to $\mathrm{cof}\mathbf{F}$, we can compose convex non-decreasing functions of linear combinations of these invariants and retain polyconvexity. Here, we propose the following general form of the strain energy function:

\begin{multline}
\Psi(\mathbf{F}) = \Psi_{I_1}(I_1)+   \Psi_{I_2}(I_2)+   \Psi_{I_{4v}}(I_{4v}) +\Psi_{I_{4w}}(I_{4w}) + \\
\sum_{j > i} \Psi_{I_i,I_j} \left(\alpha_{ij} I_i+ (1-\alpha_{ij})I_j\right)+ \Psi_{J}(J)\, 
\label{eq_polyconvex_sum_aniso}
\end{multline}

where the second to last term includes convex non-decreasing functions of all possible linear combinations of the invariants introduced so far. The weights $\alpha_{ij}$ are restricted to the interval $(0,1)$. We could include additional terms depending on the specific application, for instance, anisotropic invariants that are convex with respect to $\mathrm{cof}\mathbf{F}$, such as $I_{5v}=\mathrm{tr}(\mathrm{cof}\mathbf{C}\mathbf{V}_0)$, can be added to this framework.
 
Turning the attention back to (\ref{eq_polyconvex_sum_aniso}), each of the $\Psi_{I_i}$ and $\Psi_{I_i, I_j}$ needs to be convex non-decreasing. This is equivalent to the derivatives $d\Psi_{I_i}/d I_i$ being monotonic functions, with $d\Psi_{I_i}/d I_i \geq 0$ in the domain of $I_i$. In the next section we show how to leverage N-ODEs to generate monotonic derivative functions and thus polyconvex strain energies.

\subsection{Neural ordinary differential equations}
\label{methods_NODE}
We are interested in finding functions that create monotonic maps between inputs $\vec{x}$ and outputs $\vec{y}$. N-ODEs are a novel architecture of neural networks that generalizes some successful models, such as residual networks \cite{he2016deep}. The key idea is to replace the number of discrete layers of classical neural networks by a continuous transformation of the hidden state by a learnable function. In this sense, the concept of depth of the neural network is replaced by time. This results in the construction of an ordinary differential equation system:
\begin{equation}
    \frac{\partial \mathbf{H}(t)}{\partial t} = \vec{f}(\mathbf{H}(t),t; \vec{\theta})
\end{equation}
Here $\mathbf{H}(t)$ represents the hidden state of the neural network. The function $f(\cdot;\vec{\theta})$ is represented by a fully-connected neural network with trainable parameters $\vec{\theta}$. The relationship between the input and the output for this model can be obtained by integrating the system in an arbitrary interval of time:
\begin{equation}
    \mathbf{H}(1) = \mathbf{H}(0) + \int_0^1 \vec{f}(\mathbf{H}(t),t; \vec{\theta}) dt
\end{equation}
Here, we assign the input to the initial condition $\vec{x} = \mathbf{H}(0)$ and the output to the final state $\vec{y} = \mathbf{H}(1)$. 

We know from ordinary differential equation analysis that the solution trajectories never intersect each other in the state space, as illustrated in Fig.~\ref{fig_ODE_mapping}. Intersection of trajectories would imply that for a given point in the state space there is more than one value for the rate of change $f(\mathbf{H}(t))$, which contravenes the definition of the ordinary differential equation system. In one-dimensional case, this condition implies that for two different trajectories ${\rm H}_1(t)$ and ${\rm H}_2(t)$:
\begin{eqnarray}
{\rm H}_2(0) \geq {\rm H}_1(0) & \iff & {\rm H}_2(1) \geq {\rm H}_1(1)\\
{\rm H}_2(0) < {\rm H}_1(0) & \iff & {\rm H}_2(1) < {\rm H}_1(1)
\end{eqnarray}
These conditions can be succinctly written in terms of the input $x = {\rm H}(0)$ and output $y = {\rm H}(1)$ of the neural ordinary differential equations as:
\begin{equation}
    (y_2 - y_1)(x_2 - x_1)\geq  0 \label{eq:nondec}
\end{equation}
which correspond exactly to the requirements for a monotonic function. To satisfy the polyconvexity requirements, we also need to ensure that these functions are non-negative in the domain of the input. To achieve this property, we need to ensure that 
\begin{equation}
\int_0^1 f({\rm H}(t)) dt \geq \max\{0,-{\rm H}_{min}\}, \hspace{0.1cm} {\rm H}(0) = {\rm H}_{min}    
\end{equation}
 where ${\rm H}_{min}$ is lowest possible value of the input. Although there are multiple ways to satisfy this condition, here we focus on the simplest case, where ${\rm H}_{min} = 0$ and $f(0) = 0$. This particular scenario can be achieved by shifting the inputs and removing all the bias parameters from the neural network that approximates $f$ and setting the initial condition of the ODE as $H(0)=0$. In this case, the neural network that approximates the right-hand side of the ordinary differential equation can be written as:
 \begin{equation}
     f(x) = \mathbf{W}_n h(...\mathbf{W}_2 h(\mathbf{W}_1x)))
 \end{equation}
 where $h(\cdot)$ is non-linear activation function applied element-wise, $n$ is the depth of the network and $\mathbf{W}_i$ are the learnable parameters.
 With this setup, we have shown that the functions approximated by one-dimensional N-ODEs are monotonic and non-negative, and suitable to construct polyconvex strain energy functions.

\subsection*{Anisotropic, hyperelastic, and fully incompressible materials} 
To demonstrate the potential of the proposed framework, we will attempt to learn an anisotropic, hyperelastic and incompressible materials. We have introduced the strain energy, $\Psi$, which is a function of the right Cauchy-Green deformation tensor, $\mathbf{C}$, and two material direction vectors, $\mathbf{v}_0$ and $\mathbf{w}_0$. For incompressible materials, the last term of the strain energy, $\Psi_J$, is replaced by the constraint $p(J-1)$ with $p$ a Lagrange multiplier. Furthermore, in the case of hyperelastic materials the free energy has no other contribution. The second Piola-Kirchhoff stress tensor, $\mathbf{S}$, follows from the Doyle-Erickson formula by differentiating the strain energy $\Psi$ with respect to $\mathbf{C}$ \cite{DOYLE195653},


\begin{equation}
    \mathbf{S} = 2\frac{\partial\Psi}{\partial \mathbf{C}} = 2 \frac{\partial \Psi}{\partial I_1} \mathbf{I} + 2 \frac{\partial \Psi}{\partial I_2} (I_1 \mathbf{I} - \mathbf{C}) + 2 \frac{\partial \Psi}{\partial I_{4v}} \mathbf{V}_0 + 2 \frac{\partial \Psi}{\partial I_{4w}} \mathbf{W}_0 + p\mathbf{C}^{-1} \, .
    \label{eq_S}
\end{equation}

Often times in finite element packages, the Cauchy stress, $\mathbf{\sigma}$, is used, which can be obtained via a push forward operation. 
To solve the equilibrium equations we also need the elasticity tensor:

\begin{equation}
    \mathbb{C} = 2\frac{\partial \mathbf{S}}{\partial \mathbf{C} }.
    \label{eq_CC}
\end{equation}

The full expansion of the elasticity tensor is shown in the Appendix. Also note that if the finite element implementation is done in the deformed configuration, the spatial version of the elasticity tensor, $\mathbb{c}$, can be obtained via a push forward operation.

Therefore, given N-ODEs describing the strain energy derivatives, these functions can be used directly to determine the stress for a given deformation by evaluating (\ref{eq_S}). For nonlinear finite element simulations, evaluation of the tangent (\ref{eq_CC}) requires the second derivatives of the strain energy, which in our case involves taking derivatives of the N-ODEs with respect to their inputs.

\subsection*{Data-driven polyconvex constitutive models using neural ordinary differential equations}

We have shown that N-ODEs produce monotonic functions. We therefore use this tool to represent the derivatives of the terms in eq. (\ref{eq_polyconvex_sum_aniso}), i.e. $\partial \Psi_{I_1}/\partial I_1, \partial \Psi_{I_2}/\partial I_2$, etc. as illustrated in Fig. \ref{fig_diagram}. By producing monotonic non-negative derivative functions, we guarantee that the underlying strain energy terms are convex non-decreasing functions of the invariants, which in turn guarantees polyconvexity of the energy with respect to $\mathbf{F}$. A total of $10$ N-ODEs are used, each corresponding to a different partial derivative of the strain energy. The architecture of each of the N-ODEs is the same and summarized in Table \ref{table01}.

\begin{table}[h!]\centering
\caption{Neural network architecture}
\label{table01}
\begin{tabularx}{0.78\textwidth}{lll}
\hline
Layer           & Number of nodes   & Activation function\\ \hline
Input           & 1                 & None\\ 
Hidden layer 1  & 5                 & tanh \\ 
Hidden layer 2  & 5                 & tanh\\
Output          & 1                 & Linear\\
\hline
\end{tabularx}
\end{table}

While the monotonicity condition is guaranteed directly by using N-ODEs, to satisfy the non-negative conditions, $\partial \Psi_{I_i}/\partial I_i \geq 0$, we first note that  $I_1, I_2 \geq 3$ for an incompressible material \cite{schroder2003invariant}. For these invariants, setting $J_1=I_1 - 3, J_2=I_2 - 3$ as inputs to the N-ODE and setting biases to zero gives the initial condition  $H(0)=0$. A single non-negative bias is added after the evaluation of each of these two N-ODEs, i.e. $H_1$, $H_2$ for $\partial \Psi_{I_1}/\partial I_1 , \partial \Psi_{I_2}/\partial I_2$, resulting in $\partial \Psi_{I_1}/\partial I_1 , \partial \Psi_{I_2}/\partial I_2 \geq 0$ at $J_1=0,J_2=0$. In this way, the functions $\Psi_{I_1}, \Psi_{I_2}$ are convex non-decreasing with a minimum of $H_1,H_2$ at $I_1=3,I_2=3$, which not only guarantees the polyconvexity of the strain energy but also allows for the model to capture polyconvex models such as Mooney-Rivlin and neo-Hookean. We remark that this formulation can be extended to compressible materials using modified version of $I_1, I_2$. Namely, the usual split of the deformation gradient into isochoric and volumetric parts results in the isochoric invariant $\bar{I}_1\geq 3$, such that $\bar{I}_1$ is polyconvex with respect to $\mathbf{F}$ \cite{schroder2003invariant}. The isochoric invariant $\bar{I}_2$ is not polyconvex with respect to $\mathbf{F}$. However, dividing by an appropriate power of $J$, one can obtain modified invariants $\hat{I}_2\geq3$ which are polyconvex \cite{schroder2003invariant}. 


For the anisotropic terms, the domain of the input is $I_{4v},I_{4w} \geq 0$, with $I_{4v} = I_{4w} = 1$ for the identity map. We set shifted invariants $J_{4v}=I_{4v} -1, J_{4w}=I_{4w} - 1$ as the inputs of the N-ODEs and remove all biases to get $\partial \Psi_{I_{4v}}/\partial I_{4v},\partial \Psi_{I_{4w}}/\partial I_{4w} \geq 0$, and $\partial \Psi_{I_{4v}}/\partial I_{4v},\partial \Psi_{I_{4w}}/\partial I_{4w} = 0$ for $I_{4v},I_{4w}=1$. For the mixed terms in eq (\ref{eq_polyconvex_sum_aniso}) we also set all biases to zero, but, additionally, we check if the output of the N-ODE is non-negative otherwise it is set to zero. 

The N-ODE framework outlined here is general, it will always produce monotonic derivative functions and thus convex functions of the invariants. Specifying the minimum of the energy and the stress-free state can be achieved in variety of ways and we show one convenient solution. Note also that it is possible to incorporate anisotropic compression terms by considering the invariants $I_{5v} = \cof \mathbf{C}: \mathbf{v}_0 \otimes \mathbf{v}_0$ and $I_{5w} = \cof \mathbf{C}: \mathbf{w}_0 \otimes \mathbf{w}_0$, which are convex with respect to $\mathrm{cof}\mathbf{F}$ and represent area changes orthogonal to the fibers. Extension to compressible materials is also straightforward. Lastly, even though we focus on hyperelasticity, polyconvex strain energies can be used as building blocks to describe dissipative mechanisms.

\subsubsection*{Model calibration and verification}
\label{methods_training_data}

The training and validation data for the N-ODEs are taken from biaxial tests on porcine skin. The data corresponds to five experimental protocols described in Table \ref{table02}. In this case, the experiments are described in terms of the principal stretches of the deformation gradient:
\begin{equation}
    \mathbf{F} = \begin{bmatrix}
                   \lambda_{xx} & 0 & 0 \\ 0 & \lambda_{yy} & 0 \\ 0 & 0 & \lambda_{zz}
                 \end{bmatrix}
\end{equation}
With the assumption of plane stress incompressible behavior, $J = \lambda_{xx}\lambda_{yy}\lambda_{zz} = 1$ leads to $\lambda_{zz} = 1/\lambda_{xx} \lambda_{yy}$ plus a boundary condition to evaluate the pressure Lagrange multiplier $p$.

\begin{table}[h!]\centering
\caption{Biaxial experimental protocols. $\lambda_x$ and $\lambda_y$ represent the stretches imposed in the $x$ and $y$ directions and $\sigma_z$ is the stress in the $z$ direction. The parameter $\lambda$ is monotonically increased during the experiments to stretch the tissue. The last component of the deformation gradient is directly obtained from the incompressible constraint, while the vanishing normal stress is imposed as the necessary boundary condition to solve for the pressure $p$.}
\label{table02}
\begin{tabularx}{0.4\textwidth}{llll}
\hline
Loading         & $\lambda_{xx}$       & $\lambda_{yy}$       & $\sigma_{zz}$\\ \hline
Off-x           & $\sqrt{\lambda}$  & $\lambda$         & 0\\ 
Off-y           & $\lambda$         & $\sqrt{\lambda}$  & 0\\ 
Equibiaxial     & $\lambda$         & $\lambda$         & 0\\
Strip-x         & $\lambda$         & 1                 & 0\\
Strip-y         & 1                 & $\lambda$         & 0\\
\hline
\end{tabularx}
\end{table}

We also test the N-ODE against synthetic data generated using four popular analytical models summarized next. 

\subsubsection*{Holzapfel-Gasser-Ogden (HGO) }
The HGO material model proposed in \cite{holzapfel2000HGO} assumes there are two families of fibers that contribute to the energy through an exponential term. The strain energy is,
\begin{equation}
    \Psi (\mathbf{C}, \mathbf{v}_0, \mathbf{w}_0) = \Psi_{\text{iso}} (\mathbf{C}) + \Psi_{\text{aniso}} (\mathbf{C}, \mathbf{v}_0, \mathbf{w}_0) + p(J-1) ,
\end{equation}
with dependence on two anisotropy directions $\mathbf{v}_0,\mathbf{w}_0$, 
\begin{equation}
    \Psi_{\text{aniso}} (\mathbf{C}, \mathbf{v}_0, \mathbf{w}_0) = \frac{k_1}{2k_2} \sum_{i=4v,4w} \left\{\exp \left[k_2 (I_i - 1)^2 \right] -1 \right\} \, .
    \label{GOHpsianiso}
\end{equation}
The parameters controlling the anisotropic contribution are $k_1,k_2$. The strain invariants $I_{4v},I_{4w}$ are the same as defined in eq. (\ref{eq_invs}). The isotropic contribution depends on the first invariant, $I_1$, and is that of a neo-Hookean solid
\begin{equation}
    \Psi_{\text{iso}}(\mathbf{C}) = \mu (I_1-3)\, ,
\end{equation}
with parameter $\mu$. The incompressibility constraint is imposed through the Lagrange multiplier $p$.  

\subsubsection*{Gasser-Ogden-Holzapfel (GOH) }
The GOH material model was initially proposed  to model arterial walls \cite{gasser2005GOH}. It is an extension to the HGO model which considers a single fiber family but incorporates fiber dispersion. Since then, it has been used to model other soft tissues including skin. The strain energy density function in this model 
\begin{equation}
    \Psi (\mathbf{C}, \mathbf{v}_0) = \Psi_{\text{iso}} (\mathbf{C}) + \Psi_{\text{aniso}} (\mathbf{C}, \mathbf{v}_0) + p(J-1)\, ,
\end{equation}
where 
\begin{align}
    &\Psi_{\text{iso}}(\mathbf{C}) = \mu (I_1-3) \, ,
    \label{GOHpsiiso}
    \\
    &\Psi_{\text{aniso}} (\mathbf{C}, \mathbf{v}_0) = \frac{k_1}{4k_2} \left[exp \left(k_2 E^2 \right) -1 \right] \, ,
    \label{GOHpsianiso}
\end{align}
and
\begin{equation}
    E = \left[\kappa I_1 + (1-3\kappa) I_{4v} - 1 \right]\, .
\end{equation}

The parameters are the same as those used in the HGO model, except for the fiber dispersion parameter $\kappa$. The strain invariants, $I_1$ and $I_{4v}$ are the same as defined in (\ref{eq_invs}), with the single anisotropy direction $\mathbf{v}_0$.

\subsubsection*{Mooney Rivlin (MR)}
Originally, the MR models was proposed to capture the mechanical response of rubber-like materials in large strains \cite{Mooney1940, Rivlin1948}. There are a number of different formulations for MR models, here we use 


\begin{equation}
    \Psi(\mathbf{\lambda}) = C_{10}(I_1-3) + C_{01}(I_2-3) + C_{20}(I_1-3)^2 \, + p(J-1),
\end{equation}


parameterized by $C_{10}, C_{01}, C_{20}$.

\subsubsection*{Fung-type models }
The strain energy function proposed by Fung et al. was one of the first models of soft tissue to capture the strain-stiffening anisotropy of collagenous tissue \cite{fung1979}. Unlike the previous material models, the strain energy proposed by Fung et al. is directly in terms of the Green-Lagrange strain, $\mathbf{E}=(\mathbf{C}-\mathbf{I})/2$, 
\begin{equation}
    \Psi(\mathbf{C}) = \frac{c_1}{2}\exp(a_1 E_{xx}^2 + a_2 E_{yy}^2 + 2a_4 E_{xx}E_{yy})
    \label{eq_fung}
\end{equation}
with parameters $a_1,a_2,a_4,c_1$. Note that equation (\ref{eq_fung}) is strictly a model of a two-dimensional material and, as originally introduced in \cite{fung1979}, it cannot be used for three-dimensional elasticity problems. However, since its original development, generalized forms of the model by Fung et al. have been developed, e.g. \cite{sun2005finite}. In this work we use the original form based on \cite{fung1979}.

\subsubsection*{Finite element implementation}

A major motivation for the data-driven framework described here is to make it readily usable in large scale finite element simulations and realistic applications. Since the N-ODEs directly encode energy derivatives, it is straighforward to implement the N-ODE material model  through the UANISOHYPER subroutine in the commercial finite element package Abaqus. The implementation is available in the Github link at the end. 

\section{Results}

\begin{figure}[t]
\centering
\includegraphics[width=0.9\linewidth]{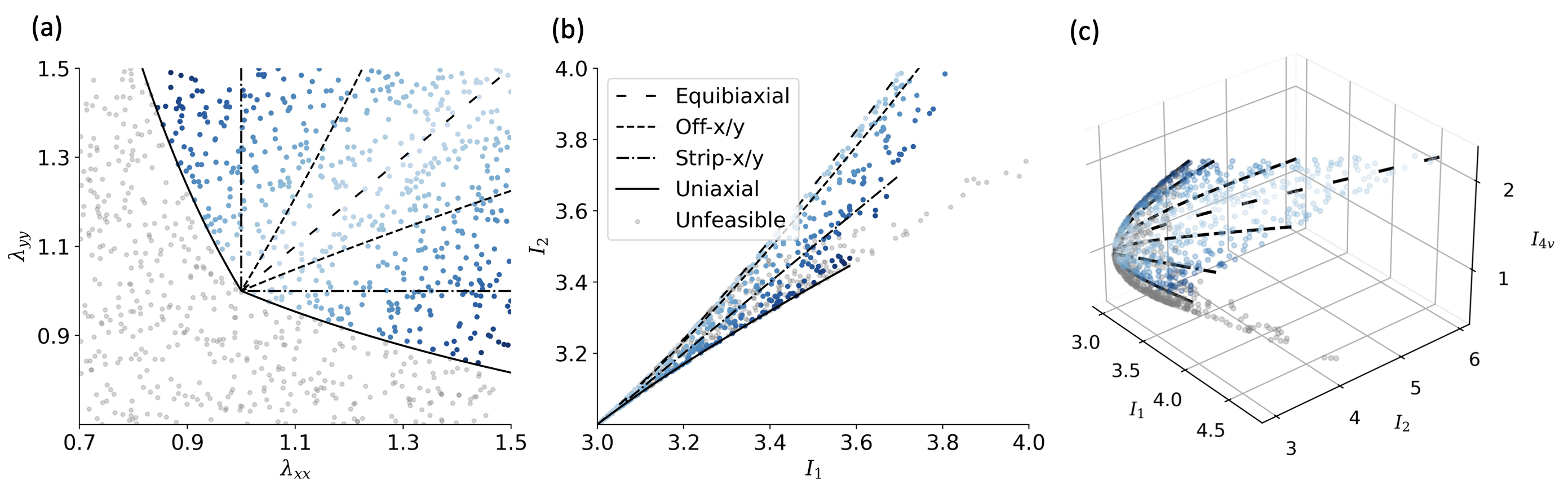}
\caption{Input space of the N-ODE in (a) $\lambda_{xx}-\lambda_{yy}$ space, $I_1-I_2$ space, and (c) $I_1-I_2-I_{4v}$ space. Points in the input space where testing is not feasible are shown with grey markers, whereas feasible points are in shades of blue. The boundary of the feasible region is marked by the uniaxial tests in $x$ and $y$ directions in $\lambda_{xx}-\lambda_{yy}$ space in (a). The map to the space of invariants is nonlinear and condenses the points into a narrow cone in the $I_1-I_2$ space in (b). Including anisotropy, the input domain for the data-driven models continues to show how evenly space points in the $\lambda_{xx}-\lambda_{yy}$ space are nonlinearly mapped to the domain of the N-ODEs.  }
\label{fig_input_space} 
\end{figure}

The model should estimate the stresses for arbitrary deformations of the material. In experimental mechanics of thin specimens such as skin, the most appropriate method to quantify material behavior is through biaxial tests in terms of the principal stretches, $\lambda_{xx},\lambda_{yy}$. However, as outlined above, the model inputs are the invariants of the right Cauchy-Green deformation tensor, $I_i$. In Fig. \ref{fig_input_space}a we show how the input space for biaxial experiments maps into the input space of the N-ODEs, Fig. \ref{fig_input_space}b,c.
The protocols in Table~\ref{table02} are shown as curves in \ref{fig_input_space}a, the scatter points in shades of blue denote deformations in the feasible region of  $\lambda_{xx}-\lambda_{yy}$ space, and the scatter points in gray denote infeasible regions. Points colored gray are not achievable during the testing of thin specimens like skin because they correspond to a compressive state under which thin membranes would buckle. The boundary between the feasible and infeasible regions are uniaxial tests in the $x$ and $y$ directions.

The same points from Fig. \ref{fig_input_space}a are mapped into the invariant space. A projection onto the space $I_1-I_2$ is depicted in Fig. \ref{fig_input_space}b, and a projection onto $I_1-I_2-I_{4v}$ is shown in Fig. \ref{fig_input_space}c. The space $I_1,I_2$ corresponds to the input space for the isotropic material behavior. The equibiaxial loading protocol falls in the middle of the feasible region in the $\lambda_{xx}-\lambda_{yy}$ space, but it is at the boundary of the $I_1-I_2$ space. The uniaxial loading cases, which determine the boundary of the feasible region in $\lambda_{xx}-\lambda_{yy}$ space also continue to bound the feasible domain in the $I_1-I_2$ space. Even though the testing points in Fig. \ref{fig_input_space}a are evenly distributed in $\lambda_{xx}-\lambda_{yy}$ space, they form a narrow cone in the $I_1-I_2$ space. The map is highly nonlinear. Since the material model is a function of the invariants, the performance of data-driven models depends on sampling the invariant space, which is not necessarily achieved by simply covering evenly spaced points in  $\lambda_{xx}-\lambda_{yy}$ space.

\begin{figure}[t]
\centering
\includegraphics[width=\linewidth]{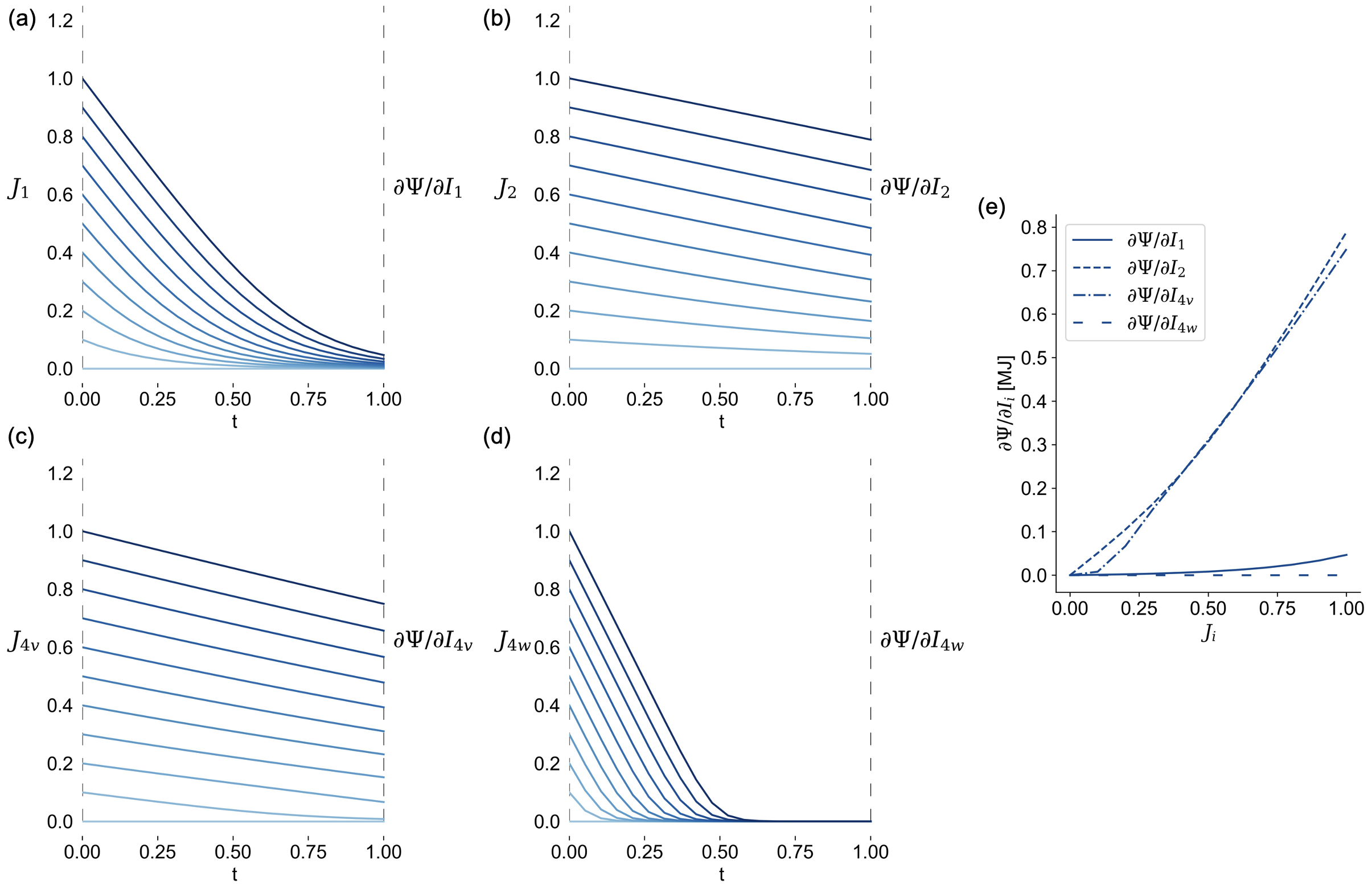}
\caption{An illustration of how the N-ODE maps points in the input space to partial derivatives of the strain energy. Time evolution of (a) $J_1$ to $\partial \Psi / \partial I_1$, (b) $J_2$ to $\partial \Psi / \partial I_2$, (c) $J_{4v}$ to $\partial \Psi / \partial I_{4v}$ and (d) $J_{4w}$ to $\partial \Psi / \partial I_{4w}$, and, (e) the shape of the resulting functions $\partial \Psi / \partial I_i$. The notation $J_i$ is used to denote shifted invariants as inputs to the N-ODE in order to satisfy the initial condition $H(0)=0$, which is one solution to achieve non-negative derivatives with a vanishing stress at the identity map.}
\label{fig_ODE_mapping} 
\end{figure}

In Fig. \ref{fig_ODE_mapping} we illustrate the principle behind the N-ODE map. Points from the input space (translated invariants $J_i$, i.e $J_1 = I_1 - 3$) are integrated in time based on the neural network, which encodes the right-hand side of an ODE. The partial derivatives of the strain energy function are defined as the solution of a N-ODE at the fictitious time $t=1$. Only the mappings for the invariants $J_1, J_2, J_{4v}$ and $J_{4w}$ are shown in Fig. \ref{fig_ODE_mapping}a-d. The monotonicity of the transformation is evident in these figures, it follows from the fact that curves originating at different initial condition never intersect as they are integrated in time. Fig. \ref{fig_ODE_mapping}e represents the direct relationship between input and output without the pseudo-time axis. 

\subsection*{Comparisons against synthetic data}

\begin{figure}[t]
\centering
\includegraphics[width=\linewidth]{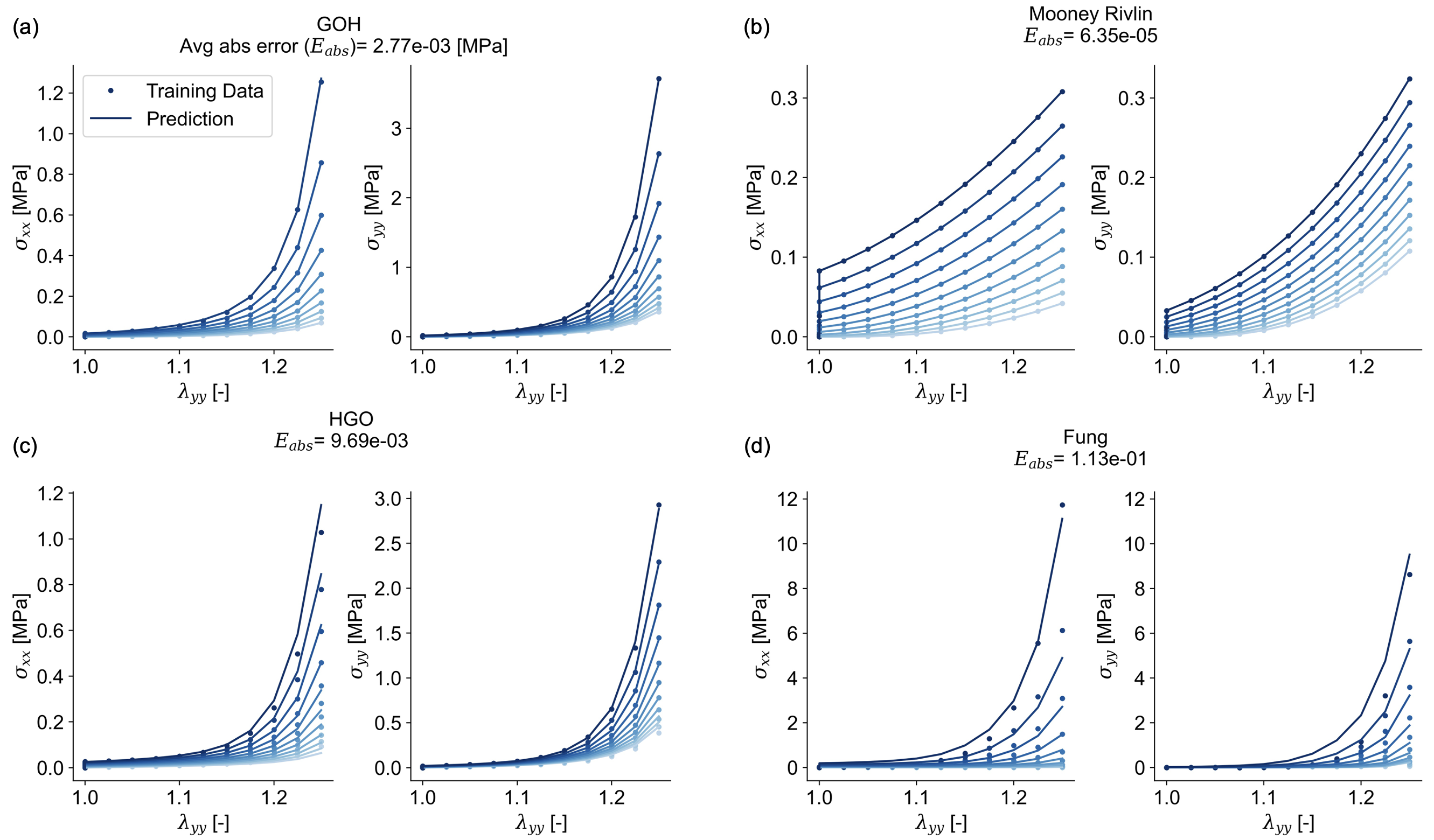}
\caption{Synthetic training data and N-ODE predictions using (a) GOH, (b) Mooney Rivlin, (c) HGO and (d) Fung type material models. The N-ODE is able to interpolate synthetic data from closed-form constitutive equations.}
\label{fig_synthetic} 
\end{figure}

We start by training the N-ODE material model with synthetic stress-stretch data generated from the four closed-form analytical models introduced in section \ref{methods_training_data}. We plot the training data as well as the predictions of the N-ODE for each of the four cases in Fig. \ref{fig_synthetic}. Even though the analytical models have each a completely different functional form, the same N-ODE architecture is capable of replicating the different synthetic datasets with high accuracy. The average absolute error is  small in all cases. The highest error occurs for the Fung model, for which stresses are an order of magnitude higher than the other models for the parameters chosen.
The best fit of the N-ODE is for the Mooney-Rivlin material. Among the four synthetic datasets, the Mooney-Rivlin is the least nonlinear. The other three analytical models include an exponential term that describes a rapid strain stiffening typical of soft tissues. Concomitantly, the exponential term in these models poses challenges for model calibration due to the extreme sensitivity to parameters \cite{lee2018}.

\begin{figure}[h!]
\centering
\includegraphics[width=\linewidth]{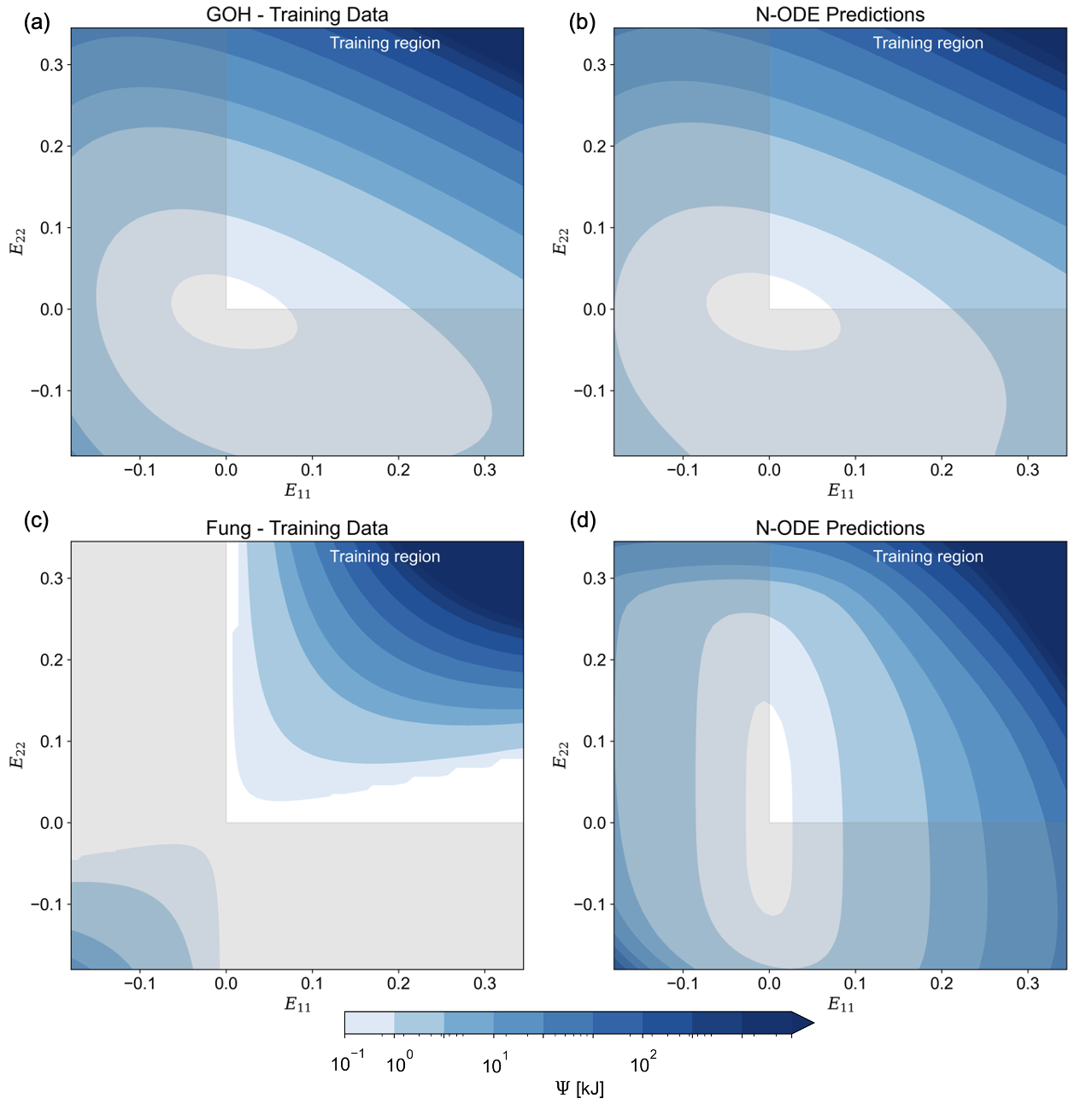}
\caption{Strain energy contours for synthetic data generated using (a) GOH and (c) Fung type material models and (b), (d) the predictions of the N-ODE after training on these datasets, respectively. The training region corresponded to tensile biaxial deformations, as indicated, while the shaded region was not included during learning. When fed non-convex data, the N-ODE still produces a polyconvex output because it is built in the formulation. }
\label{fig_convexity} 
\end{figure}
Next, we demonstrate how the N-ODE automatically satisfies polyconvexity of the strain energy. We train the N-ODE material model with synthetic data from GOH and Fung models. We show in Figs. \ref{fig_convexity}a,c the predicted contours of strain energy of the two analytical models and the corresponding N-ODE approximations are in Figs. \ref{fig_convexity}b,d. GOH is an inherently convex model \cite{gasser2005GOH}, which is reflected in the contour of the energy as a function of the Green Lagrange strain components for parameters $\mu = 0.0102, k_1 = 0.513, k_2 = 59.1, \kappa = 0.271, \theta = 1.57.$ (Fig. \ref{fig_convexity}a). Naturally, the N-ODE trained with this model also predicts a convex function (Fig. \ref{fig_convexity}b). The Fung model, on the other hand, is known to exhibit non-convex behavior for some choice of parameters, for instance $c_1 = 0.00241, a_1 = -1.75, a_2 = -21.5, a_4 = 49.8$. The N-ODE model cannot interpolate these data, it can only find a convex approximation as seen in Fig. \ref{fig_convexity}d. 

\subsection*{Performance on experimental data}

The primary objective of developing N-ODE material models is being able to capture experimental data without the restrictions of closed-form energy functions. The experimental data for this example consists of data from five biaxial tests as described in Table \ref{table02}, performed on porcine skin specimens. We divide the data in two sets, train the data on three of the biaxial test data and validate against the remaining biaxial test data. 

The N-ODE reproduces the training data almost identically, with an average error of 0.011 MPa in the worst case scenario. This is expected from the results on the synthetic datasets. The error in the validation tests is naturally higher than in the training set, but it is still in a reasonable range with average errors 1.242 MPa for off-x and 0.648 MPa for off-y data. 

\begin{figure}[h!]
\centering
\includegraphics[width=0.9\linewidth]{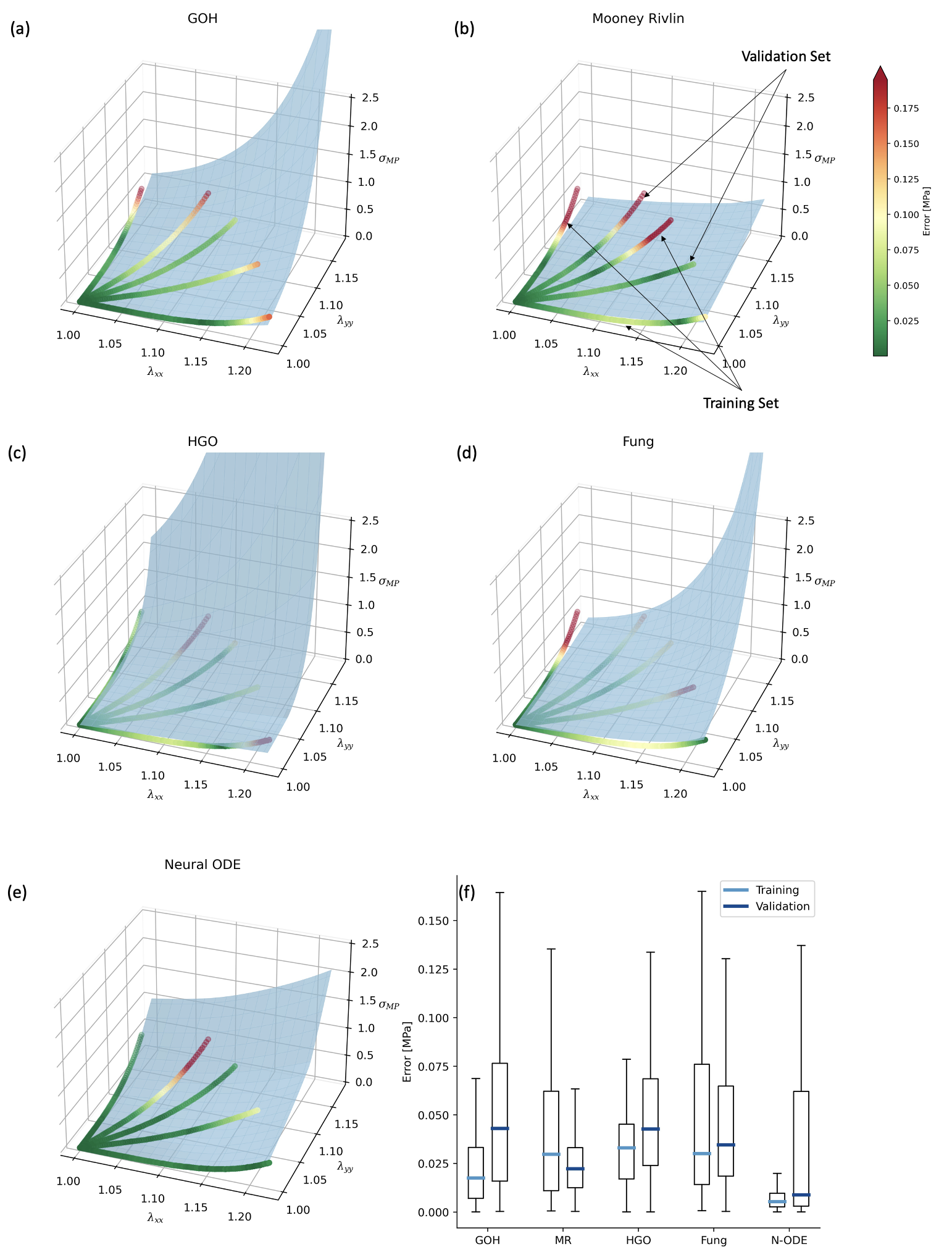}
\caption{Surfaces of maximum principal stress ($\sigma_{MP}$) and curves with the experimental data colored based on error for (a) GOH, (b) Mooney Rivlin, (c) HGO, (d) Fung and (e) N-ODE. Box plots of error distribution for the five models (f).}
\label{fig_exp2} 
\end{figure}

In Fig. \ref{fig_exp2}, we train both the N-ODE and the four analytical models with data from a different porcine skin specimen subjected to same loading as described in Table \ref{table02}. Similarly to before, we split the data into training and validation sets. The maximum principal stress predictions, $\sigma_{mp}$, from each of the analytical models as well as the N-ODE model are plotted as the graph of a surface over the $\lambda_xx,\lambda_yy$ plane (Fig. \ref{fig_exp2}a-e). The experimental data are plotted as curves colored based on the absolute error between these curves and the predicted response surface. Fig. \ref{fig_exp2}f shows a box plot of the error distribution for each model. None of the analytical models are capable of capturing the behavior of the material in the training set, whereas the N-ODE replicates the behavior in that region flawlessly. Errors naturally increase overall in the validation set. Yet, even in the validation set, the N-ODE has  the lowest median errors. Note also that under high equibiaxial stretches, the stresses in GOH, HGO and Fung models increase to unreasonably high values. This is due to the exponential form of these analytical models. The N-ODE, on the other hand, maintains convexity but does not infer an exponential growth beyond the training region. 

\begin{table}[h!]\centering
\caption{Mean absolute error, in MPa, for the validation set for the 4 analytical models and the N-ODE. Training/fitting is performed with the first 80\% of data points of each loading protocol while the last 20\% are held for validation.}
\label{table_val}
\begin{tabularx}{0.71\textwidth}{llllll}
\hline
Protocol        & GOH   & MR    & HGO   & Fung   & N-ODE \\ \hline
Off-x           & \textbf{0.083}	& 0.114	& 0.245	& 0.196	 & 0.084 \\
Off-y           & \textbf{0.010}	& 0.024	& 0.071	& 0.071	 & 0.027 \\ 
Equibiaxial     & 0.037	& 0.100	& 0.141	& 0.082	 & \textbf{0.030} \\
Strip-x         & 0.056	& 0.123	& 0.051	& 0.130	 & \textbf{0.039} \\
Strip-y         & 0.120	& 0.095	& 0.051	& 0.210	 & \textbf{0.050} \\ \hline
Average         & 0.062	& 0.091	& 0.112	& 0.138	 & \textbf{0.046} \\
\hline
\end{tabularx}
\end{table}

Table \ref{table_val} summarizes the performances of the five models from another series of tests. In each row of Table \ref{table_val} we train the models on the first 80\% of the loading path and test against the remaining 20\%. The average error from the validation test is shown in the Table. In other words, these tests are done to measure the ability of each of the models to extrapolate away from the training region, toward larger deformations. The last row in Table \ref{table_val} contains the average of the errors among the five individual tests. The N-ODE has the lowest error in three of the five cases, and is a close second in the Off-x case. The N-ODE has the lowest average error over the five tests, at 0.046 MPa. The second best performing model, GOH, has 35\% higher error relative to the N-ODE model. 

\begin{figure}[h!]
\centering
\includegraphics[width=\linewidth]{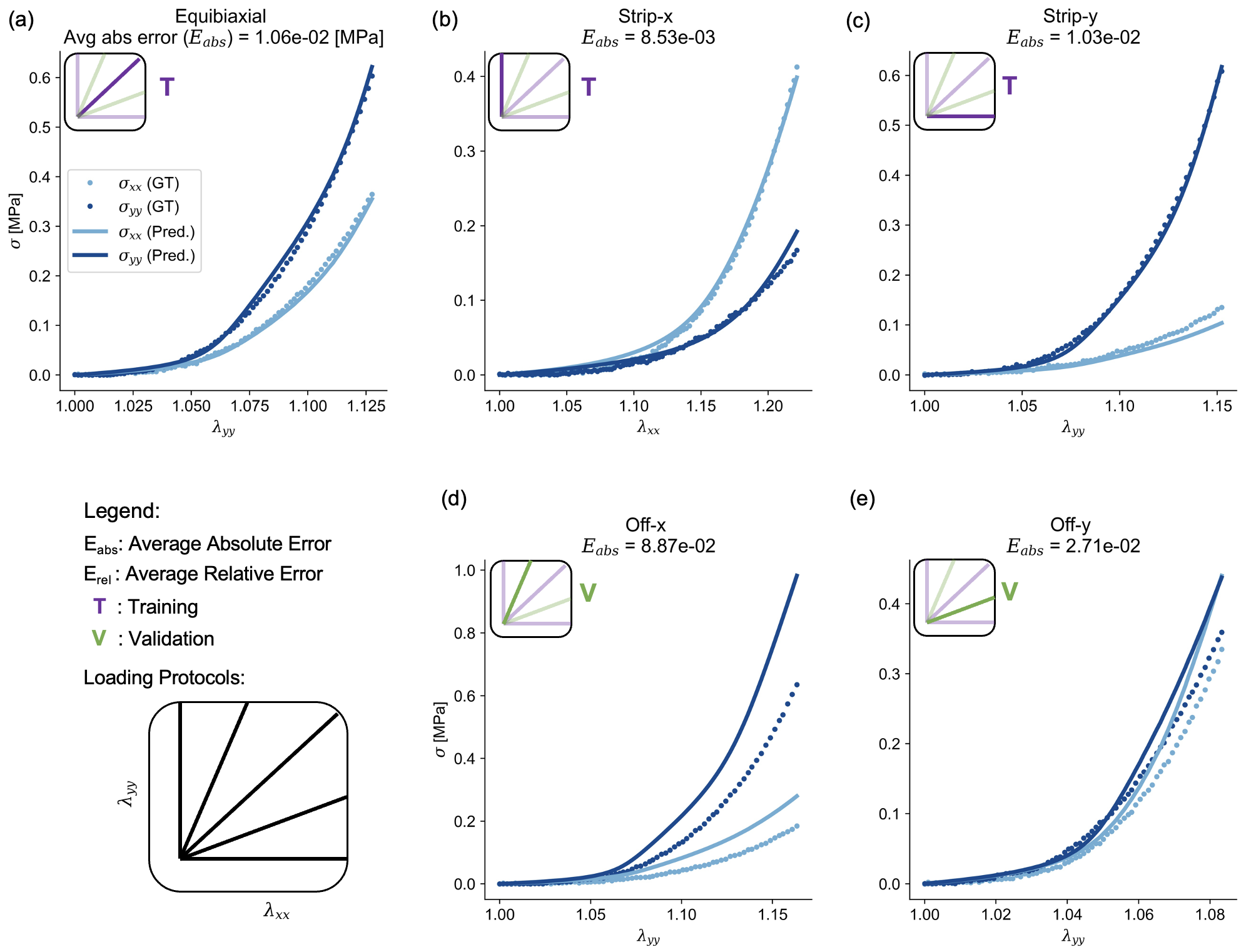}
\caption{Predictions of the N-ODE material model after training on porcine data. Test results from (a) equibiaxial, (b) Strip-x and (c) Strip-y testing protocols were used for training while (d) Off-x and (e) Off-y test results were reserved for validation.}
\label{fig_exp1} 
\end{figure}
 
\subsection*{Finite element simulations}

The results from finite element simulations in Abaqus for a range of geometries and loading scenarios using the N-ODE material model are showcased here.

\begin{figure}[h!]
\centering
\includegraphics[width=1.0\linewidth]{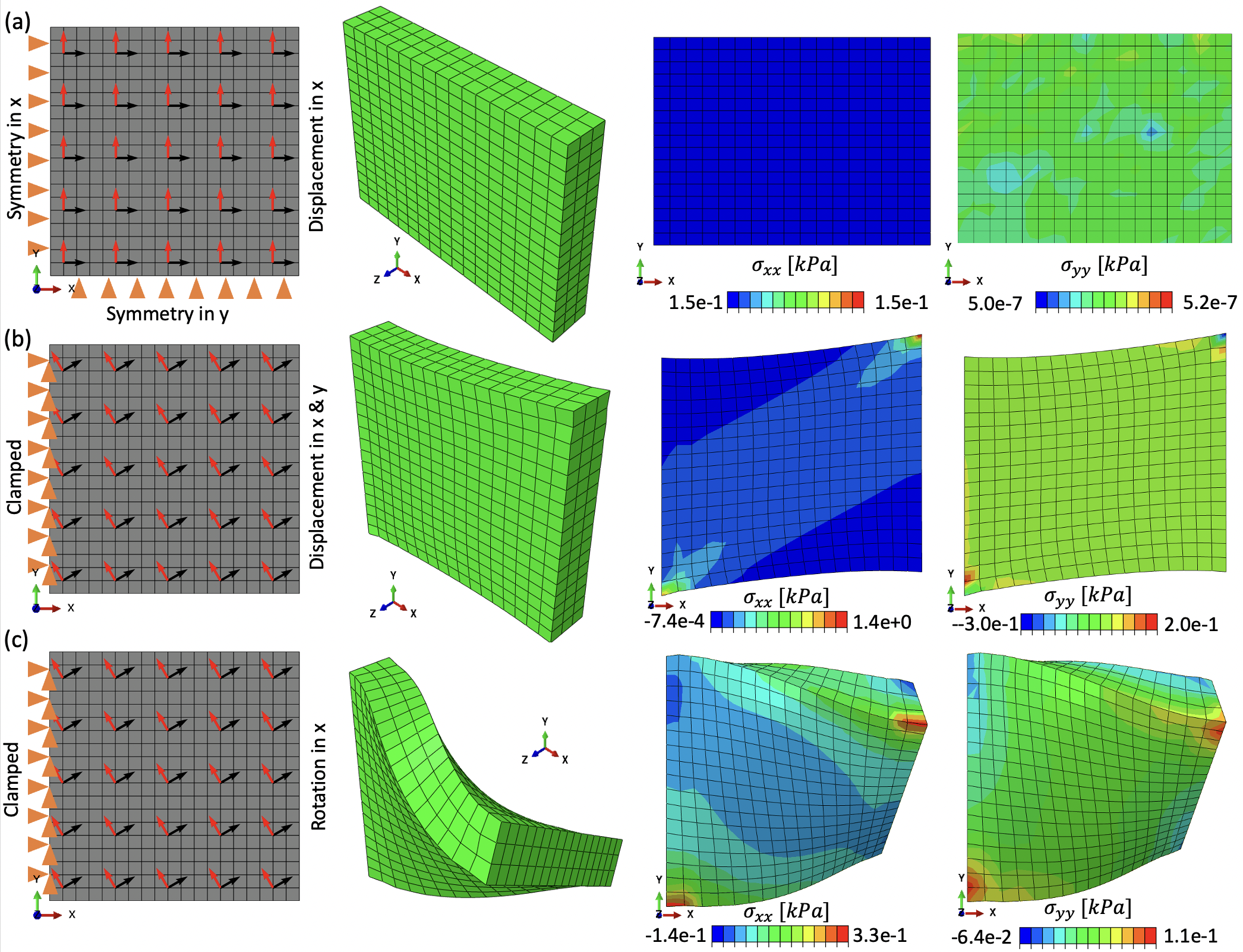}
\caption{Finite element simulations using the N-ODE material model. Boundary conditions on the undeformed geometry are shown in the first column, deformed geometry in the second column, and contours of $\sigma_{xx}$ and $\sigma_{yy}$ on the deformed geometry are depicted in the last two columns for (a) uniaxial tension, (b) shearing and stretching, (c) torsional loading scenarios. The material is anisotropic, with the directions of anisotropy depicted as quiver plots in the reference geometries. The red vector field corresponds to the stiffer orientation $\mathbf{v}_0$, while the black vector field corresponds to $\mathbf{w}_0$.}
\label{fig_FEM} 
\end{figure}

First, three basic loading scenarios were applied to a 5 $\times$ 5 cm specimen as shown in Fig.  \ref{fig_FEM} For a simple uniaxial tension test in the $x$ direction, unconstrained in $y$ the resulting stresses match the analytical evaluation (see Fig. (\ref{fig_exp1}), indicating that the subroutine is functioning as desired (Fig.\ref{fig_FEM}a).

In Fig. \ref{fig_FEM}b the left side of the specimen is held fixed while the right side is displaced both in x and y directions, whereas in \ref{fig_FEM}c a torsional loading is applied. For the shearing deformation we see a band of stress for the $\sigma_{xx}$ component, and concentration of the $\sigma_{yy}$ component at the corners but otherwise fairly uniform and small $\sigma_{yy}$ stresses since the upper and bottom surfaces are unconstrained. When a torque is applied in Fig. \ref{fig_FEM}c, stress concentrations develop in the corners where there is the most deformations as expected from the clamping condition on the boundary. In all cases the simulations converged quadratically,  demonstrating that the N-ODE model can be used as a constitutive equation in stable, nonlinear finite element simulations. 

\begin{figure}[h!]
\centering
\includegraphics[width=\linewidth]{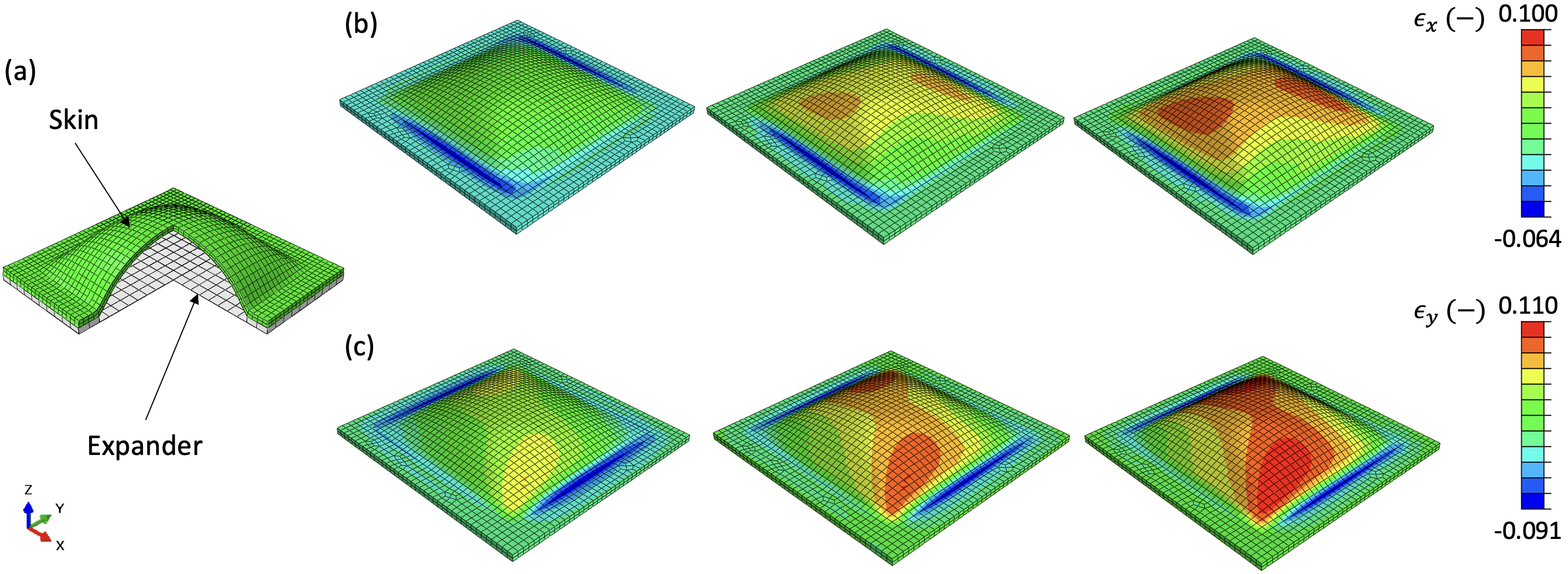}
\caption{Finite element simulations of tissue expansion using the N-ODE material model in UANISOHYPER. Model setup (a), and contours of strain (b and c) on the deformed geometry after the expander is inflated to 20, 40 and 60 cm\textsuperscript{3}, respectively.}
\label{fig_FEMexp} 
\end{figure}

Next, we performed tissue expansion simulations on $10 \times 10 \times 0.3$ cm\textsuperscript{3} a patch of skin (Fig.\ref{fig_FEMexp}a). A rectangular expander of dimensions $8 \times 8$ cm\textsuperscript{2} underneath the skin is modeled by the fluid cavity feature in Abaqus. The expander is inflated to 20, 30 and 40 cm\textsuperscript{3} resulting in the strain contours shown in Fig. \ref{fig_FEM}b and c. 

\begin{figure}[h!]
\centering
\includegraphics[width=1.0\linewidth]{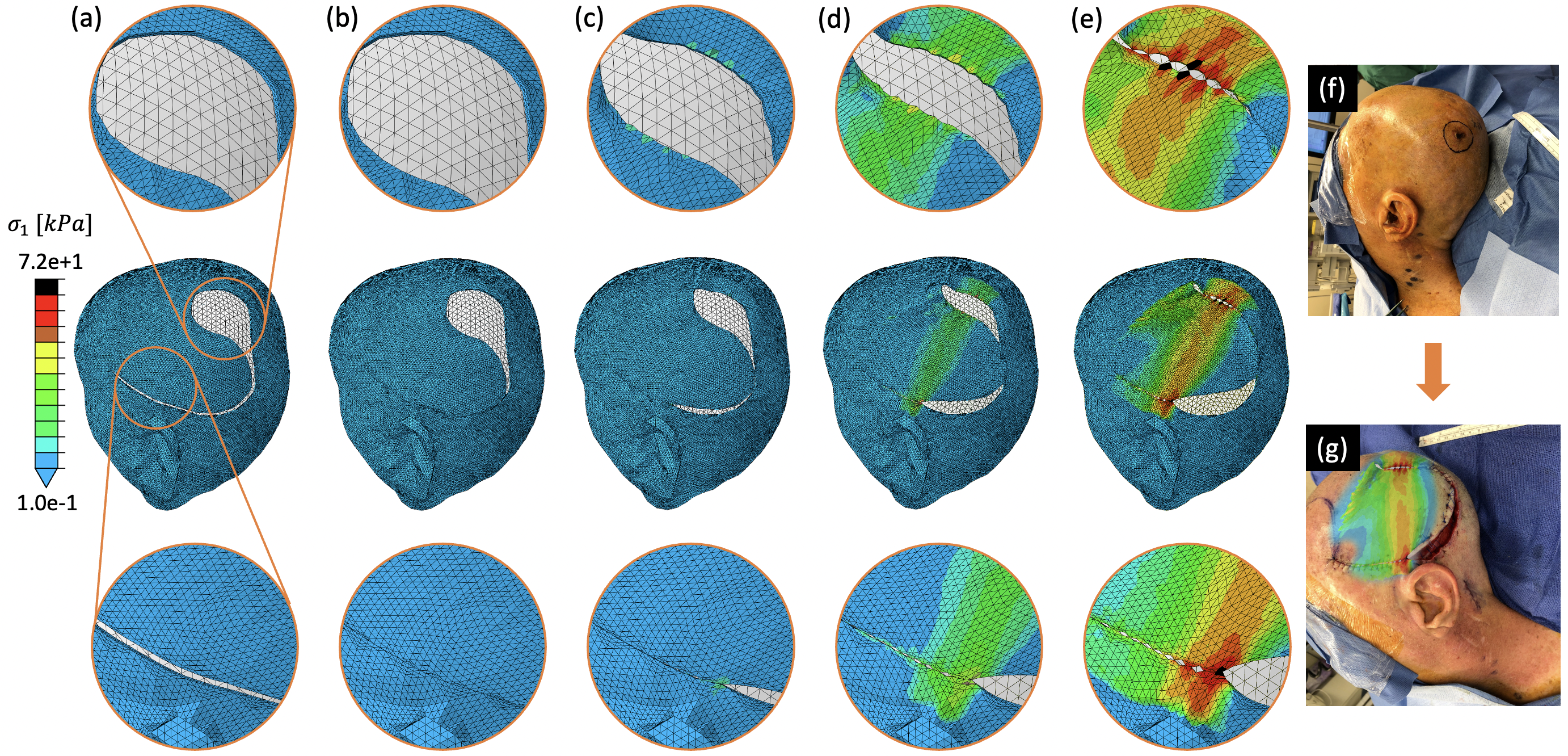}
\caption{Virtual surgery finite element model performed in Abaqus using the N-ODE material model through the user defined UANISOHYPER subroutine. Initial geometry of the scalp (a), contours of maximum principal stress on the deformed geometry after the lower sutures are completed (b), and as the upper sutures are stretched to close the wound by 30\%, 60\% and 100\% (c-e), respectively. Photographs of the scalp of the patient before the surgery (f) and after the surgery superimposed with contours of maximum principal stress (g).}
\label{fig_scalp2} 
\end{figure}

Lastly, a surgical operation was simulated on the scalp of a cancer patient with a patient specific geometry reported in \cite{lee2021personalized}. Figs.\ref{fig_scalp2} a-e show the  model during various stages of surgery. Fig.\ref{fig_scalp2}a shows the initial geometry of the model, Fig.\ref{fig_scalp2}b-e show contours of maximum principal stress on the deformed geometry when the sutures near the ear are completed, and when the sutures on top of the cranium are brought together. Fig.\ref{fig_scalp2}f shows a photo of the patient's scalp prior to surgery and Fig.\ref{fig_scalp2}g a post-operative photo of the same region with the predicted contours of maximum principal stress  superimposed.

\section{Discussion}

In this paper we present the first automatically polyconvex data-driven framework for constitutive modeling of nonlinear anisotropic materials. No other data-driven material model in the literature has been able to guarantee polyconvexity \textit{a priori}. Instead most proposed methods either do not impose convexity or rely on additional loss terms to penalize deviation from convexity during model training \cite{VLASSIS2020113299,tac2021datadriven}. 


Our approach is general enough to approximate a large class of material models. Although there might be polyconvex strain energy functions that are not captured by our solution, they might not be needed in practice. In general, our approach can represent any degree of non-linearity for a given invariant and reproduce certain types of interactions between invariants. We could arbitrarily increase the complexity of the model by adding second order interactions (i.e. $I_i + I_j + I_k$) and by composing the resulting strain energy function with other convex non-decreasing functions. If we do this multiple times, we could create deep networks to approximate an even larger class of strain energy functions. However, from what we have seen in our experiments, the capacity of the functions presented in this work is enough to model the behavior of highly nonlinear and anisotropic materials, e.g. soft tissues. 

The proposed framework is based on the principal invariants of the right Cauchy Green deformation tensor, which is commonplace in the development of anisotropic polyconvex strain energies \cite{chagnon2015hyperelastic}. However, it could be reformulated for a different set of basic invariants such as the principal stretches \cite{shariff2011physical}, or the invariants of the right stretch tensor \cite{steigmann2003isotropic,steigmann2003frame}. Specifically for anisotropy, we have restricted our attention to the pseudo-invariants $I_{4v}$ and $I_{4w}$, which are polyconvex functions of the deformation that measure the square of the stretch along the directions $\mathbf{v}_0$  and $\mathbf{w}_0$. Other anisotropic pseudo-invariants are possible. For example, as discussed in the Methods section, $I_{5v}=\mathrm{cof}\mathbf{C}:\mathbf{v}_0\otimes \mathbf{v}_0$ and $I_{5w}=\mathrm{cof}\mathbf{C}:\mathbf{w}_0\otimes \mathbf{w}_0$ are also polyconvex functions that can be used to capture energy associated with compression of the anisotropy directions \cite{schroder2003invariant}. We did not consider terms such as $I_{5v},I_{5w}$ because our data consisted of biaxial tests and because it is common to assume that fibers in soft tissues do not contribute to the energy under compression \cite{holzapfel2000HGO}. Incorporating additional directions of anisotropy, for instance based on some fiber probability distribution \cite{lanir1979structural,sacks2000biaxial}, would be straightforward. The results with only two directions were already satisfactory and we did not include additional anisotropic pseudo-invariants for that reason. 


Alternative conditions of convexity have been considered in the literature. Among them, convexity of the strain energy with respect to $\mathbf{C}$ is often sought in constitutive models \cite{VLASSIS2020113299,Holzapfel2009heart}. However, the conditions under which convexity with respect to $\mathbf{C}$ guarantees global minimizers to boundary value problems in nonlinear elasticity is not fully elucidated, see for example the recent work by Gao et al. \cite{gao2017convexity}. As pointed out in \cite{sivaloganathan2018uniqueness}, convexity with respect to $\mathbf{C}$ and polyconvexity with respect to $\mathbf{F}$ are not equivalent. Thus, in this work we focused on the condition of polyconvexity established by Ball \cite{ball1976convexity}. 

Although the proposed approach works excellently for the cases we tested, we acknowledge that there are alternatives to arrive at similar results. For example, any invertible neural network architecture should work because a function that posses an inverse in one-dimensions is monotonic. N-ODEs are an elegant and efficient solution to this problem, because the gradient can be computed with a system of ODEs rather than back-propagation, which can be expensive for a deep neural network. We also note that there are other architectures to guarantee convexity of functions based on similar principles  \cite{amos2017input}. We think our method has two main advantages: first, one can directly incorporate domain knowledge by selecting the terms and interactions that are relevant rather than using all possible inputs (strain invariants and combinations). This approach also adds interpretability, as each of the fitted functions can be inspected and compared to existing models. Second, by approximating derivatives of the strain energy function rather than the strain energy function itself, we decrease the degree of differentiability of the neural network and can employ simpler activation functions, such as ReLU. Alternative methods to deal with the issues of differentiability are integrable neural networks or regularization terms during model training \cite{VLASSIS2020113299,teichert2019machine}. Even though we do not have direct access to the strain energy function, the fact that the derivative functions are all one-dimensional enables the use of standard quadrature rules to efficiently integrate the strain energy if needed. 


Lastly, we remark that in this paper we have focused on incompressible, and hyperelastic behavior. This is not a limitation of the framework but simply the application explored in detail here. The framework produces polyconvex strain energies applicable to a much wider set of problems. For example, consideration of compressible behavior would require a suitable volumetric strain energy function which can also be data-driven and convex. Modeling of viscoelastic materials or elastoplastic deformations can also be formulated using data-driven polyconvex strain energies \cite{bonet2015computational,krishnan2014polyconvex,nordsletten2021viscoelastic}.   

\section*{Conclusions}
We present a data-driven framework to construct automatically polyconvex strain energy functions. The formulation is based on N-ODEs. We showcase the framework by using it to model the nonlinear, anisotropic, hyperelastic, and incompressible behavior of skin. The data-driven framework outperforms closed-form constitutive models. Our results foment the appeal of data-driven methods to better capture experimental material response without the constraints of analytical expressions for the strain energy while still satisfying the basic requirements of physically realistic models, i.e. the notion of polyconvexity. Additionally, the formulation is invariant and allows for the efficient computation of the second derivatives of the strain energy, which further enables the use of our data-driven framework in finite element simulations. We therefore anticipate that this work will further cement the use of data-driven methods in computational mechanics.

\section*{Acknowledgements}

This work was supported by the National Institute of Arthritis and Musculoskeletal and Skin Diseases of the National Institute of Health under award R01AR074525.

\section*{Supplementary material}
Code associated with this publication is available at a Github repository
\url{https://github.com/tajtac/NODE_v2}

%





\begin{thebibliography}{49}
\expandafter\ifx\csname natexlab\endcsname\relax\def\natexlab#1{#1}\fi
\providecommand{\bibinfo}[2]{#2}
\ifx\xfnm\relax \def\xfnm[#1]{\unskip,\space#1}\fi
\bibitem[{Ma et~al.(2017)Ma, Wu, Zhu, Xu, and Kong}]{Ma2017diagnosis}
\bibinfo{author}{J.~Ma}, \bibinfo{author}{F.~Wu}, \bibinfo{author}{J.~Zhu},
  \bibinfo{author}{D.~Xu}, \bibinfo{author}{D.~Kong},
\newblock \bibinfo{title}{A pre-trained convolutional neural network based
  method for thyroid nodule diagnosis},
\newblock \bibinfo{journal}{Ultrasonics} \bibinfo{volume}{73}
  (\bibinfo{year}{2017}) \bibinfo{pages}{221--230}.
\bibitem[{Guhathakurta(2006)}]{guhathakurta2006}
\bibinfo{author}{P.~Guhathakurta},
\newblock \bibinfo{title}{Long-range monsoon rainfall prediction of 2005 for
  the districts and sub-division kerala with artificial neural network},
\newblock \bibinfo{journal}{Current Science} \bibinfo{volume}{90}
  (\bibinfo{year}{2006}) \bibinfo{pages}{773--779}.
\bibitem[{Casey et~al.(2020)Casey, Son, Bilionis, and
  Barnes}]{casey2020prediction}
\bibinfo{author}{A.~D. Casey}, \bibinfo{author}{S.~F. Son},
  \bibinfo{author}{I.~Bilionis}, \bibinfo{author}{B.~C. Barnes},
\newblock \bibinfo{title}{Prediction of energetic material properties from
  electronic structure using 3d convolutional neural networks},
\newblock \bibinfo{journal}{Journal of Chemical Information and Modeling}
  \bibinfo{volume}{60} (\bibinfo{year}{2020}) \bibinfo{pages}{4457--4473}.
\bibitem[{Duraisamy et~al.(2019)Duraisamy, Iaccarino, and Xiao}]{Durasaimy2019}
\bibinfo{author}{K.~Duraisamy}, \bibinfo{author}{G.~Iaccarino},
  \bibinfo{author}{H.~Xiao},
\newblock \bibinfo{title}{Turbulence modeling in the age of data},
\newblock \bibinfo{journal}{Annual Review of Fluid Mechanics}
  \bibinfo{volume}{51} (\bibinfo{year}{2019}) \bibinfo{pages}{357--377}.
\bibitem[{Lee et~al.(1999)Lee, Almond, and Harris}]{lee1999fatigue}
\bibinfo{author}{J.~Lee}, \bibinfo{author}{D.~Almond},
  \bibinfo{author}{B.~Harris},
\newblock \bibinfo{title}{The use of neural networks for the prediction of
  fatigue lives of composite materials},
\newblock \bibinfo{journal}{Composites Part A: Applied Science and
  Manufacturing} \bibinfo{volume}{30} (\bibinfo{year}{1999})
  \bibinfo{pages}{1159--1169}.
\bibitem[{Lu et~al.(2020)Lu, Dao, Kumar, Ramamurty, Karniadakis, and
  Suresh}]{lu2020multifidelity}
\bibinfo{author}{L.~Lu}, \bibinfo{author}{M.~Dao}, \bibinfo{author}{P.~Kumar},
  \bibinfo{author}{U.~Ramamurty}, \bibinfo{author}{G.~E. Karniadakis},
  \bibinfo{author}{S.~Suresh},
\newblock \bibinfo{title}{Extraction of mechanical properties of materials
  through deep learning from instrumented indentation},
\newblock \bibinfo{journal}{Proceedings of the National Academy of Sciences of
  the United States of America} \bibinfo{volume}{117} (\bibinfo{year}{2020})
  \bibinfo{pages}{7052--7062}.
\bibitem[{Wang et~al.(2021)Wang, Huan, and Garikipati}]{wang2021variational}
\bibinfo{author}{Z.~Wang}, \bibinfo{author}{X.~Huan},
  \bibinfo{author}{K.~Garikipati},
\newblock \bibinfo{title}{Variational system identification of the partial
  differential equations governing microstructure evolution in materials:
  Inference over sparse and spatially unrelated data},
\newblock \bibinfo{journal}{Computer Methods in Applied Mechanics and
  Engineering} \bibinfo{volume}{377} (\bibinfo{year}{2021})
  \bibinfo{pages}{113706}.
\bibitem[{Le et~al.(2015)Le, Yvonnet, and He}]{Le2015}
\bibinfo{author}{B.~A. Le}, \bibinfo{author}{J.~Yvonnet},
  \bibinfo{author}{Q.-C. He},
\newblock \bibinfo{title}{Computational homogenization of nonlinear elastic
  materials using neural networks},
\newblock \bibinfo{journal}{International Journal for Numerical Methods in
  Engineering} \bibinfo{volume}{104} (\bibinfo{year}{2015})
  \bibinfo{pages}{1061--1084}.
\bibitem[{Liu et~al.(2020)Liu, Liang, and Sun}]{liu2020}
\bibinfo{author}{M.~Liu}, \bibinfo{author}{L.~Liang}, \bibinfo{author}{W.~Sun},
\newblock \bibinfo{title}{A generic physics-informed neural network-based
  constitutive model for soft biological tissues},
\newblock \bibinfo{journal}{Computer Methods in Applied Mechanics and
  Engineering} \bibinfo{volume}{372} (\bibinfo{year}{2020}).
\bibitem[{Peng et~al.(2020)Peng, Alber, Tepole, Cannon, De, Dura-Bernal,
  Garikipati, Karniadakis, Lytton, Perdikaris et~al.}]{peng2020multiscale}
\bibinfo{author}{G.~C. Peng}, \bibinfo{author}{M.~Alber},
  \bibinfo{author}{A.~B. Tepole}, \bibinfo{author}{W.~R. Cannon},
  \bibinfo{author}{S.~De}, \bibinfo{author}{S.~Dura-Bernal},
  \bibinfo{author}{K.~Garikipati}, \bibinfo{author}{G.~Karniadakis},
  \bibinfo{author}{W.~W. Lytton}, \bibinfo{author}{P.~Perdikaris}, et~al.,
\newblock \bibinfo{title}{Multiscale modeling meets machine learning: What can
  we learn?},
\newblock \bibinfo{journal}{Archives of Computational Methods in Engineering}
  (\bibinfo{year}{2020}) \bibinfo{pages}{1--21}.
\bibitem[{Zhang and Garikipati(2020)}]{garikipati2020multiresolution}
\bibinfo{author}{X.~Zhang}, \bibinfo{author}{K.~Garikipati},
\newblock \bibinfo{title}{Machine learning materials physics: Multi-resolution
  neural networks learn the free energy and nonlinear elastic response of
  evolving microstructures},
\newblock \bibinfo{journal}{Computer Methods in Applied Mechanics and
  Engineering} \bibinfo{volume}{372} (\bibinfo{year}{2020})
  \bibinfo{pages}{113362}.
\bibitem[{Tac et~al.(2021)Tac, Sree, Rausch, and Tepole}]{tac2021datadriven}
\bibinfo{author}{V.~Tac}, \bibinfo{author}{V.~D. Sree}, \bibinfo{author}{M.~K.
  Rausch}, \bibinfo{author}{A.~B. Tepole}, \bibinfo{title}{Data-driven modeling
  of the mechanical behavior of anisotropic soft biological tissue},
  \bibinfo{year}{2021}.
\bibitem[{Marsden and Hughes(1994)}]{marsden1994mathematical}
\bibinfo{author}{J.~E. Marsden}, \bibinfo{author}{T.~J. Hughes},
  \bibinfo{title}{Mathematical foundations of elasticity},
  \bibinfo{publisher}{Courier Corporation}, \bibinfo{year}{1994}.
\bibitem[{Kuhl et~al.(2006)Kuhl, Askes, and Steinmann}]{kuhl2006illustration}
\bibinfo{author}{E.~Kuhl}, \bibinfo{author}{H.~Askes},
  \bibinfo{author}{P.~Steinmann},
\newblock \bibinfo{title}{An illustration of the equivalence of the loss of
  ellipticity conditions in spatial and material settings of hyperelasticity},
\newblock \bibinfo{journal}{European Journal of Mechanics-A/Solids}
  \bibinfo{volume}{25} (\bibinfo{year}{2006}) \bibinfo{pages}{199--214}.
\bibitem[{Ball(1976)}]{ball1976convexity}
\bibinfo{author}{J.~M. Ball},
\newblock \bibinfo{title}{Convexity conditions and existence theorems in
  nonlinear elasticity},
\newblock \bibinfo{journal}{Archive for rational mechanics and Analysis}
  \bibinfo{volume}{63} (\bibinfo{year}{1976}) \bibinfo{pages}{337--403}.
\bibitem[{Gasser et~al.(2005)Gasser, Ogden, and Holzapfel}]{gasser2005GOH}
\bibinfo{author}{T.~C. Gasser}, \bibinfo{author}{R.~W. Ogden},
  \bibinfo{author}{G.~A. Holzapfel},
\newblock \bibinfo{title}{Hyperelastic modelling of arterial layers with
  distributed collagen fibre orientations},
\newblock \bibinfo{journal}{Journal of the royal society interface}
  \bibinfo{volume}{3} (\bibinfo{year}{2005}) \bibinfo{pages}{15--35}.
\bibitem[{Holzapfel et~al.(2000)Holzapfel, Gasser, and
  Ogden}]{holzapfel2000HGO}
\bibinfo{author}{G.~A. Holzapfel}, \bibinfo{author}{T.~C. Gasser},
  \bibinfo{author}{R.~W. Ogden},
\newblock \bibinfo{title}{A new constitutive framework for arterial wall
  mechanics and a comparative study of material models},
\newblock \bibinfo{journal}{Journal of Elasticity} \bibinfo{volume}{61}
  (\bibinfo{year}{2000}) \bibinfo{pages}{1--48}.
\bibitem[{Fung et~al.(1979)Fung, Fronek, and Patitucci}]{fung1979}
\bibinfo{author}{Y.~C. Fung}, \bibinfo{author}{F.~Fronek},
  \bibinfo{author}{P.~Patitucci},
\newblock \bibinfo{title}{Pseudoelasticity of arteries and the choice of its
  mathematical expression},
\newblock \bibinfo{journal}{American Journal of Physiology}
  \bibinfo{volume}{237} (\bibinfo{year}{1979}) \bibinfo{pages}{H620--H631}.
\bibitem[{Chagnon et~al.(2015)Chagnon, Rebouah, and
  Favier}]{chagnon2015hyperelastic}
\bibinfo{author}{G.~Chagnon}, \bibinfo{author}{M.~Rebouah},
  \bibinfo{author}{D.~Favier},
\newblock \bibinfo{title}{Hyperelastic energy densities for soft biological
  tissues: a review},
\newblock \bibinfo{journal}{Journal of Elasticity} \bibinfo{volume}{120}
  (\bibinfo{year}{2015}) \bibinfo{pages}{129--160}.
\bibitem[{Ehret and Itskov(2007)}]{ehret2007polyconvex}
\bibinfo{author}{A.~E. Ehret}, \bibinfo{author}{M.~Itskov},
\newblock \bibinfo{title}{A polyconvex hyperelastic model for fiber-reinforced
  materials in application to soft tissues},
\newblock \bibinfo{journal}{Journal of Materials Science} \bibinfo{volume}{42}
  (\bibinfo{year}{2007}) \bibinfo{pages}{8853--8863}.
\bibitem[{Limbert(2019)}]{limbert2019skin}
\bibinfo{author}{G.~Limbert}, \bibinfo{title}{Skin Biophysics: From
  Experimental Characterisation to Advanced Modelling},
  volume~\bibinfo{volume}{22}, \bibinfo{publisher}{Springer},
  \bibinfo{year}{2019}.
\bibitem[{Jor et~al.(2013)Jor, Parker, Taberner, Nash, and
  Nielsen}]{jor2013computational}
\bibinfo{author}{J.~W. Jor}, \bibinfo{author}{M.~D. Parker},
  \bibinfo{author}{A.~J. Taberner}, \bibinfo{author}{M.~P. Nash},
  \bibinfo{author}{P.~M. Nielsen},
\newblock \bibinfo{title}{Computational and experimental characterization of
  skin mechanics: identifying current challenges and future directions},
\newblock \bibinfo{journal}{Wiley Interdisciplinary Reviews: Systems Biology
  and Medicine} \bibinfo{volume}{5} (\bibinfo{year}{2013})
  \bibinfo{pages}{539--556}.
\bibitem[{Mueller et~al.(2021)Mueller, Elrod, Distler, Schiestl, and
  Mazza}]{mueller2021reliability}
\bibinfo{author}{B.~Mueller}, \bibinfo{author}{J.~Elrod},
  \bibinfo{author}{O.~Distler}, \bibinfo{author}{C.~Schiestl},
  \bibinfo{author}{E.~Mazza},
\newblock \bibinfo{title}{On the reliability of suction measurements for skin
  characterization},
\newblock \bibinfo{journal}{Journal of Biomechanical Engineering}
  \bibinfo{volume}{143} (\bibinfo{year}{2021}) \bibinfo{pages}{021002}.
\bibitem[{Lee et~al.(2018)Lee, Turin, Gosain, Bilionis, and Tepole}]{lee2018}
\bibinfo{author}{T.~Lee}, \bibinfo{author}{S.~Y. Turin}, \bibinfo{author}{A.~K.
  Gosain}, \bibinfo{author}{I.~Bilionis}, \bibinfo{author}{A.~B. Tepole},
\newblock \bibinfo{title}{Propagation of material behavior uncertainty in a
  nonlinear finite element model of reconstructive surgery},
\newblock \bibinfo{journal}{Biomechanics and Modeling in Mechanobiology}
  \bibinfo{volume}{17} (\bibinfo{year}{2018}) \bibinfo{pages}{1857--1873}.
\bibitem[{Leng et~al.(2021)Leng, Calve, and Tepole}]{leng2021}
\bibinfo{author}{Y.~Leng}, \bibinfo{author}{S.~Calve}, \bibinfo{author}{A.~B.
  Tepole},
\newblock \bibinfo{title}{Predicting the mechanical properties of fibrin using
  neural networks trained on discrete fiber network data},
\newblock \bibinfo{journal}{arXiv preprint arXiv:2101.11712}
  (\bibinfo{year}{2021}).
\bibitem[{Vlassis and Sun(2021)}]{Vlassis202elastoplast}
\bibinfo{author}{N.~N. Vlassis}, \bibinfo{author}{W.~Sun},
\newblock \bibinfo{title}{Sobolev training of thermodynamic-informed neural
  networks for interpretable elasto-plasticity models with level set
  hardening},
\newblock \bibinfo{journal}{Computer Methods in Applied Mechanics and
  Engineering} \bibinfo{volume}{377} (\bibinfo{year}{2021}).
\bibitem[{Chen et~al.(2019)Chen, Rubanova, Bettencourt, and
  Duvenaud}]{chen2019node}
\bibinfo{author}{R.~T.~Q. Chen}, \bibinfo{author}{Y.~Rubanova},
  \bibinfo{author}{J.~Bettencourt}, \bibinfo{author}{D.~Duvenaud},
  \bibinfo{title}{Neural ordinary differential equations},
  \bibinfo{year}{2019}.
\bibitem[{Schr{\"o}der(2010)}]{schroder2010anisotropic}
\bibinfo{author}{J.~Schr{\"o}der},
\newblock \bibinfo{title}{Anisotropic polyconvex energies},
\newblock in: \bibinfo{booktitle}{Poly-, quasi-and rank-one convexity in
  applied mechanics}, \bibinfo{publisher}{Springer}, \bibinfo{year}{2010}, pp.
  \bibinfo{pages}{53--105}.
\bibitem[{Schr{\"o}der and Neff(2003)}]{schroder2003invariant}
\bibinfo{author}{J.~Schr{\"o}der}, \bibinfo{author}{P.~Neff},
\newblock \bibinfo{title}{Invariant formulation of hyperelastic transverse
  isotropy based on polyconvex free energy functions},
\newblock \bibinfo{journal}{International journal of solids and structures}
  \bibinfo{volume}{40} (\bibinfo{year}{2003}) \bibinfo{pages}{401--445}.
\bibitem[{He et~al.(2016)He, Zhang, Ren, and Sun}]{he2016deep}
\bibinfo{author}{K.~He}, \bibinfo{author}{X.~Zhang}, \bibinfo{author}{S.~Ren},
  \bibinfo{author}{J.~Sun},
\newblock \bibinfo{title}{Deep residual learning for image recognition},
\newblock in: \bibinfo{booktitle}{Proceedings of the IEEE conference on
  computer vision and pattern recognition}, pp. \bibinfo{pages}{770--778}.
\bibitem[{Doyle and Ericksen(1956)}]{DOYLE195653}
\bibinfo{author}{T.~Doyle}, \bibinfo{author}{J.~Ericksen},
\newblock \bibinfo{title}{Nonlinear elasticity},
\newblock volume~\bibinfo{volume}{4} of \textit{\bibinfo{series}{Advances in
  Applied Mechanics}}, \bibinfo{publisher}{Elsevier}, \bibinfo{year}{1956}, pp.
  \bibinfo{pages}{53--115}.
\bibitem[{Mooney(1940)}]{Mooney1940}
\bibinfo{author}{M.~Mooney},
\newblock \bibinfo{title}{A theory of large elastic deformation},
\newblock \bibinfo{journal}{Journal of Applied Physics} \bibinfo{volume}{11}
  (\bibinfo{year}{1940}).
\bibitem[{Rivlin(1948)}]{Rivlin1948}
\bibinfo{author}{R.~S. Rivlin},
\newblock \bibinfo{title}{Large elastic deformations of isotropic materials iv.
  further developments of the general theory},
\newblock \bibinfo{journal}{Philosophical Transactions of the Royal Society A}
  \bibinfo{volume}{241} (\bibinfo{year}{1948}).
\bibitem[{Sun and Sacks(2005)}]{sun2005finite}
\bibinfo{author}{W.~Sun}, \bibinfo{author}{M.~S. Sacks},
\newblock \bibinfo{title}{Finite element implementation of a generalized
  fung-elastic constitutive model for planar soft tissues},
\newblock \bibinfo{journal}{Biomechanics and modeling in mechanobiology}
  \bibinfo{volume}{4} (\bibinfo{year}{2005}) \bibinfo{pages}{190--199}.
\bibitem[{Lee et~al.(2021)Lee, Turin, Stowers, Gosain, and
  Tepole}]{lee2021personalized}
\bibinfo{author}{T.~Lee}, \bibinfo{author}{S.~Y. Turin},
  \bibinfo{author}{C.~Stowers}, \bibinfo{author}{A.~K. Gosain},
  \bibinfo{author}{A.~B. Tepole},
\newblock \bibinfo{title}{Personalized computational models of
  tissue-rearrangement in the scalp predict the mechanical stress signature of
  rotation flaps},
\newblock \bibinfo{journal}{The Cleft Palate-Craniofacial Journal}
  \bibinfo{volume}{58} (\bibinfo{year}{2021}) \bibinfo{pages}{438--445}.
\bibitem[{Vlassis et~al.(2020)Vlassis, Ma, and Sun}]{VLASSIS2020113299}
\bibinfo{author}{N.~N. Vlassis}, \bibinfo{author}{R.~Ma},
  \bibinfo{author}{W.~Sun},
\newblock \bibinfo{title}{Geometric deep learning for computational mechanics
  part i: anisotropic hyperelasticity},
\newblock \bibinfo{journal}{Computer Methods in Applied Mechanics and
  Engineering} \bibinfo{volume}{371} (\bibinfo{year}{2020})
  \bibinfo{pages}{113299}.
\bibitem[{Shariff(2011)}]{shariff2011physical}
\bibinfo{author}{M.~Shariff},
\newblock \bibinfo{title}{Physical invariants for nonlinear orthotropic
  solids},
\newblock \bibinfo{journal}{International journal of solids and structures}
  \bibinfo{volume}{48} (\bibinfo{year}{2011}) \bibinfo{pages}{1906--1914}.
\bibitem[{Steigmann(2003{\natexlab{a}})}]{steigmann2003isotropic}
\bibinfo{author}{D.~J. Steigmann},
\newblock \bibinfo{title}{On isotropic, frame-invariant, polyconvex
  strain-energy functions},
\newblock \bibinfo{journal}{Quarterly Journal of Mechanics and Applied
  Mathematics} \bibinfo{volume}{56} (\bibinfo{year}{2003}{\natexlab{a}})
  \bibinfo{pages}{483--491}.
\bibitem[{Steigmann(2003{\natexlab{b}})}]{steigmann2003frame}
\bibinfo{author}{D.~J. Steigmann},
\newblock \bibinfo{title}{Frame-invariant polyconvex strain-energy functions
  for some anisotropic solids},
\newblock \bibinfo{journal}{Mathematics and mechanics of Solids}
  \bibinfo{volume}{8} (\bibinfo{year}{2003}{\natexlab{b}})
  \bibinfo{pages}{497--506}.
\bibitem[{Lanir(1979)}]{lanir1979structural}
\bibinfo{author}{Y.~Lanir},
\newblock \bibinfo{title}{A structural theory for the homogeneous biaxial
  stress-strain relationships in flat collagenous tissues},
\newblock \bibinfo{journal}{Journal of biomechanics} \bibinfo{volume}{12}
  (\bibinfo{year}{1979}) \bibinfo{pages}{423--436}.
\bibitem[{Sacks(2000)}]{sacks2000biaxial}
\bibinfo{author}{M.~S. Sacks},
\newblock \bibinfo{title}{Biaxial mechanical evaluation of planar biological
  materials},
\newblock \bibinfo{journal}{Journal of elasticity and the physical science of
  solids} \bibinfo{volume}{61} (\bibinfo{year}{2000})
  \bibinfo{pages}{199--246}.
\bibitem[{Holzapfel and Ogden(2009)}]{Holzapfel2009heart}
\bibinfo{author}{G.~A. Holzapfel}, \bibinfo{author}{R.~W. Ogden},
\newblock \bibinfo{title}{{Constitutive modelling of passive myocardium: a
  structurally based framework for material characterization.}},
\newblock \bibinfo{journal}{Philosophical transactions. Series A, Mathematical,
  physical, and engineering sciences} \bibinfo{volume}{367}
  (\bibinfo{year}{2009}) \bibinfo{pages}{3445--75}.
\bibitem[{Gao et~al.(2017)Gao, Neff, Roventa, and Thiel}]{gao2017convexity}
\bibinfo{author}{D.~Y. Gao}, \bibinfo{author}{P.~Neff},
  \bibinfo{author}{I.~Roventa}, \bibinfo{author}{C.~Thiel},
\newblock \bibinfo{title}{On the convexity of nonlinear elastic energies in the
  right cauchy-green tensor},
\newblock \bibinfo{journal}{Journal of Elasticity} \bibinfo{volume}{127}
  (\bibinfo{year}{2017}) \bibinfo{pages}{303--308}.
\bibitem[{Sivaloganathan and Spector(2018)}]{sivaloganathan2018uniqueness}
\bibinfo{author}{J.~Sivaloganathan}, \bibinfo{author}{S.~J. Spector},
\newblock \bibinfo{title}{On the uniqueness of energy minimizers in finite
  elasticity},
\newblock \bibinfo{journal}{Journal of Elasticity} \bibinfo{volume}{133}
  (\bibinfo{year}{2018}) \bibinfo{pages}{73--103}.
\bibitem[{Amos et~al.(2017)Amos, Xu, and Kolter}]{amos2017input}
\bibinfo{author}{B.~Amos}, \bibinfo{author}{L.~Xu}, \bibinfo{author}{J.~Z.
  Kolter},
\newblock \bibinfo{title}{Input convex neural networks},
\newblock in: \bibinfo{booktitle}{International Conference on Machine
  Learning}, \bibinfo{organization}{PMLR}, pp. \bibinfo{pages}{146--155}.
\bibitem[{Teichert et~al.(2019)Teichert, Natarajan, Van~der Ven, and
  Garikipati}]{teichert2019machine}
\bibinfo{author}{G.~H. Teichert}, \bibinfo{author}{A.~Natarajan},
  \bibinfo{author}{A.~Van~der Ven}, \bibinfo{author}{K.~Garikipati},
\newblock \bibinfo{title}{Machine learning materials physics: Integrable deep
  neural networks enable scale bridging by learning free energy functions},
\newblock \bibinfo{journal}{Computer Methods in Applied Mechanics and
  Engineering} \bibinfo{volume}{353} (\bibinfo{year}{2019})
  \bibinfo{pages}{201--216}.
\bibitem[{Bonet et~al.(2015)Bonet, Gil, and Ortigosa}]{bonet2015computational}
\bibinfo{author}{J.~Bonet}, \bibinfo{author}{A.~J. Gil},
  \bibinfo{author}{R.~Ortigosa},
\newblock \bibinfo{title}{A computational framework for polyconvex large strain
  elasticity},
\newblock \bibinfo{journal}{Computer Methods in Applied Mechanics and
  Engineering} \bibinfo{volume}{283} (\bibinfo{year}{2015})
  \bibinfo{pages}{1061--1094}.
\bibitem[{Krishnan and Steigmann(2014)}]{krishnan2014polyconvex}
\bibinfo{author}{J.~Krishnan}, \bibinfo{author}{D.~J. Steigmann},
\newblock \bibinfo{title}{A polyconvex formulation of isotropic
  elastoplasticity theory},
\newblock \bibinfo{journal}{IMA Journal of Applied Mathematics}
  \bibinfo{volume}{79} (\bibinfo{year}{2014}) \bibinfo{pages}{722--738}.
\bibitem[{Nordsletten et~al.(2021)Nordsletten, Capilnasiu, Zhang, Wittgenstein,
  Hadjicharalambous, Sommer, Sinkus, and
  Holzapfel}]{nordsletten2021viscoelastic}
\bibinfo{author}{D.~Nordsletten}, \bibinfo{author}{A.~Capilnasiu},
  \bibinfo{author}{W.~Zhang}, \bibinfo{author}{A.~Wittgenstein},
  \bibinfo{author}{M.~Hadjicharalambous}, \bibinfo{author}{G.~Sommer},
  \bibinfo{author}{R.~Sinkus}, \bibinfo{author}{G.~A. Holzapfel},
\newblock \bibinfo{title}{A viscoelastic model for human myocardium},
\newblock \bibinfo{journal}{arXiv preprint arXiv:2105.06671}
  (\bibinfo{year}{2021}).

\end{thebibliography}


\appendix
\section{Explicit expression for the 4\textsuperscript{th} order elasticity tensor in the reference configuration}
\label{appendixA}
The explicit expression for the 4\textsuperscript{th} order elasticity tensor for the plane stress problem is given here. In the following equations 2D tensors are denoted with the subscript "s". Note, this is not the standard derivative of the second Piola Kirchhoff stress tensor with respect to the Cauchy Green deformation tensor. This is because with the incompressible and plane stress assumptions, the out of plane stretch and the pressure are a functino of the in-plane deformations. These dependencies are taken into account for the computation of the consistent tangent, 

\begin{multline}
    \mathbb{C}_s = 2\frac{\mathbf{S}_s}{\mathbf{C}_s} = 
    \delta_1 \mathbf{I}_s \otimes \mathbf{I}_s
    + 
    \delta_2 (\mathbf{C}_s^{-1} \otimes \mathbf{I}_s+\mathbf{I}_s \otimes \mathbf{C}_s^{-1})
    +
    \delta_3 (\mathbf{C}_s \otimes \mathbf{I}_s+\mathbf{I}_s \otimes \mathbf{C}_s) + \\
    \delta_4 (\mathbf{C}_s \otimes \mathbf{C}_s^{-1} + \mathbf{C}_s^{-1} \otimes \mathbf{C}_s)
    + 
    \delta_5 \mathbf{C}_s \otimes \mathbf{C}_s 
    +
    \delta_6 \mathbb{I}_s  
    +  
    \delta_7 \mathbf{V}_{0s} \otimes \mathbf{V}_{0s}
    + 
    \delta_8 \mathbf{W}_{0s} \otimes \mathbf{W}_{0s} 
    +
    \delta_9 \mathbf{C}_s^{-1} \otimes \mathbf{C}_s^{-1} 
    + \\
    \delta_{10} \mathbf{C}_s^{-1} \odot \mathbf{C}_s^{-1} 
    + 
    \delta_{11} (\mathbf{I}_s \otimes \mathbf{V}_{0s} + \mathbf{V}_{0s} \otimes \mathbf{I}_s)
    + 
    \delta_{12} (\mathbf{I}_s \otimes \mathbf{W}_{0s} + \mathbf{W}_{0s} \otimes \mathbf{I}_s)
    + \\
    \delta_{13} (\mathbf{C}_s^{-1} \otimes \mathbf{V}_{0s} + \mathbf{V}_{0s} \otimes \mathbf{C}_s^{-1})
    + 
    \delta_{14} (\mathbf{C}_s^{-1} \otimes \mathbf{W}_{0s} + \mathbf{W}_{0s} \otimes \mathbf{C}_s^{-1})
    +
    \delta_{15} (\mathbf{C}_s \otimes \mathbf{V}_{0s} + \mathbf{V}_{0s} \otimes \mathbf{C}_s)
    + \\
    \delta_{16} (\mathbf{C}_s \otimes \mathbf{W}_{0s} + \mathbf{W}_{0s} \otimes \mathbf{C}_s)
    \, ,
\end{multline}
where
\begin{align*}
    \delta_1 &= \hphantom{-}4 \left[ \frac{\partial^2 \Psi}{\partial I_1^2} + 2\frac{\partial^2 \Psi}{\partial I_1 \partial I_2}(I_1 + C_{33}) + \frac{\partial^2 \Psi}{\partial I_2^2} (I_1^2+2I_1C_{33}+C_{33}^2) + \frac{\partial \Psi}{\partial I_2} \right] 
    \\
    \delta_2 &= \hphantom{-}4 \left[-\frac{\partial^2 \Psi}{\partial I_1^2} C_{33} - \frac{\partial^2 \Psi}{\partial I_1 \partial I_2}C_{33}(2I_1+C_{33}) - \frac{\partial^2 \Psi }{\partial I_2^2} I_1 C_{33}(I_1+C_{33}) - \frac{\partial \Psi}{\partial I_2} C_{33} \right] 
    \\
    \delta_3 &= \hphantom{-}4 \left[ - \frac{\partial^2 \Psi}{\partial I_1 \partial I_2} - \frac{\partial^2 \Psi}{\partial I_2^2} (I_1 + C_{33}) \right]
    \\
    \delta_4 &= \hphantom{-}4 \left[ \frac{\partial^2 \Psi}{\partial I_1 \partial I_2} C_{33} +  \frac{\partial^2 \Psi}{\partial I_2^2}I_1C_{33} \right]
    \\
    \delta_5 &= \hphantom{-}4 \frac{\partial^2 \Psi}{\partial I_2^2} 
    \\
    \delta_6 &= \hphantom{-}4 \frac{\partial \Psi}{\partial I_2}  
    \\
    \delta_7 &= \hphantom{-}4 \frac{\partial^2 \Psi}{\partial I_{4v}^2}
    \\
    \delta_8 &= \hphantom{-}4 \frac{\partial^2 \Psi}{\partial I_{4w}^2}
    \\
    \delta_9 &= \hphantom{-}4 \left[\frac{\partial^2 \Psi}{\partial I_1^2} C_{33}^2 + 2\frac{\partial^2 \Psi}{\partial I_1 \partial I_2}I_1C_{33}^2 + \frac{\partial \Psi}{\partial I_1} C_{33} + \frac{\partial^2 \Psi}{\partial I_2^2} I_1^2 C_{33}^2 + \frac{\partial \Psi}{\partial I_2} I_1 C_{33} \right] 
    \\
    \delta_{10} &= \hphantom{-}4\left[ \frac{\partial \Psi}{\partial I_1} C_{33} + \frac{\partial \Psi}{\partial I_2} I_1 C_{33} \right] 
    \\
    \delta_{11} &= \hphantom{-}4 \left[ \frac{\partial^2 \Psi}{\partial I_1 \partial I_{4v}} + \frac{\partial^2 \Psi}{\partial I_2 \partial I_{4v}} (I_1 + C_{33}) \right]
    \\
    \delta_{12} &= \hphantom{-}4 \left[ \frac{\partial^2 \Psi}{\partial I_1 \partial I_{4w}} + \frac{\partial^2 \Psi}{\partial I_2 \partial I_{4w}} (I_1 + C_{33})  \right]
    \\
    \delta_{13} &= \hphantom{-}4 \left[ -\frac{\partial^2 \Psi}{\partial I_1 \partial I_{4v}}C_{33} - \frac{\partial^2 \Psi}{\partial I_2 \partial I_{4v}} I_1 C_{33} \right]
    \\
    \delta_{14} &= \hphantom{-}4 \left[ -\frac{\partial^2 \Psi}{\partial I_1 \partial I_{4w}}C_{33} - \frac{\partial^2 \Psi}{\partial I_2 \partial I_{4w}} I_1 C_{33} \right]
    \\
    \delta_{15} &= -4 \frac{\partial^2 \Psi}{\partial I_2 \partial I_{4v}}
    \\
    \delta_{16} &= -4 \frac{\partial^2 \Psi}{\partial I_2 \partial I_{4w}}
    \, ,
\end{align*}
$\mathbb{I}$ is the 4\textsuperscript{th} order identity tensor, $C_{33} = 1/\det{\mathbf{C}}$ and the special product noted by $(\odot)$ is defined as $(\bullet \odot \circ)_{ijkl} = [(\bullet) _{ik} (\circ) _{jl} + (\bullet) _{il} (\circ) _{jk}]/2$.

\end{document}